\documentclass{article}

\usepackage[utf8]{inputenc} %

\usepackage{microtype}
\usepackage{graphicx}
\usepackage{subfigure}
\usepackage{booktabs} %
\usepackage{float}
\usepackage{dblfloatfix}
\usepackage{hyperref}

\usepackage[accepted]{icml2023}

\usepackage{amsmath}
\usepackage{amssymb}
\usepackage{mathtools}
\usepackage{amsthm}

\usepackage[capitalize,noabbrev]{cleveref}

\theoremstyle{plain}

\theoremstyle{definition}

\theoremstyle{remark}

\usepackage{xcolor}
\usepackage{soul}
\usepackage{cleveref}

\newcommand{\remove}[1]{}

\usepackage{xspace,xpunctuate}
\newcommand{\ie}{{i.e.,}\xspace}
\newcommand{\eg}{{e.g.,}\xspace}
\newcommand{\ea}{{et~al\xperiod}\xspace}

\icmltitlerunning{Designing Data: Proactive Data Collection and Iteration for Machine Learning}

\begin{document}

\twocolumn[
\icmltitle{Designing Data: Proactive Data Collection and Iteration for Machine Learning Using Reflexive Planning, Monitoring, and Density Estimation}

\icmlsetsymbol{equal}{*}

\begin{icmlauthorlist}
\icmlauthor{Aspen Hopkins}{yyy,comp}
\icmlauthor{Fred Hohman}{comp}
\icmlauthor{Luca Zappella}{comp}
\icmlauthor{Xavier Suau Cuadros}{comp}
\icmlauthor{Dominik Moritz}{comp}
\end{icmlauthorlist}

\icmlaffiliation{yyy}{ Massachusetts Institute of Technology}
\icmlaffiliation{comp}{Apple}

\icmlcorrespondingauthor{Aspen Hopkins}{dataspen@mit.edu}

\icmlkeywords{Machine Learning, ICML}

\vskip 0.3in
]

\printAffiliationsAndNotice{} %

\begin{figure*}
\centering
\includegraphics[width=.7\linewidth]
  {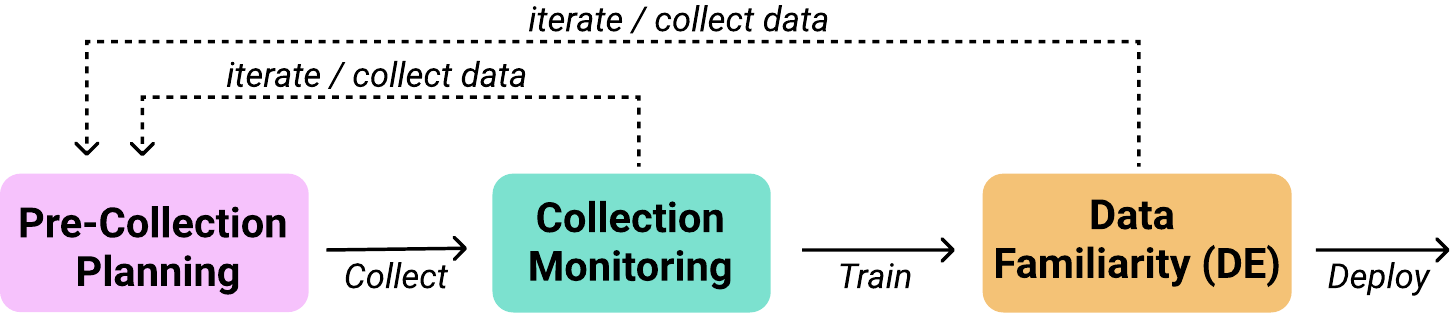}
  \caption{
\textit{Designing data} includes (I) \textbf{Pre-Collection Planning}, (II) \textbf{Collection Monitoring}, \& (III)\textbf{ Data Familiarity}. Each supplements the other, and should be used iteratively.
\vspace{-.5cm}
  }
  ~\label{fig:designing_data}
\end{figure*}
\begin{abstract}
Lack of diversity in data collection has caused significant failures in machine learning (ML) applications.
While ML developers perform post-collection interventions, these are time intensive and rarely comprehensive.
Thus, new methods to track \& manage data collection, iteration, and model training are necessary for evaluating whether datasets reflect real world variability.
We present \textit{designing data}, an iterative approach to data collection connecting HCI concepts with ML techniques.
Our process includes (1) Pre-Collection Planning, to reflexively prompt and document expected data distributions; (2) Collection Monitoring, to systematically encourage sampling diversity; and (3) Data Familiarity, to identify samples that are unfamiliar to a model using density estimation.
We apply designing data to a data collection and modeling task. 
We find models trained on ``designed'' datasets generalize better across intersectional groups than those trained on similarly sized but less targeted datasets, and that data familiarity is effective for debugging datasets.
\end{abstract}

\section{Introduction}
Curating representative training and testing datasets is fundamental to developing robust, generalizable machine learning (ML) models.
However, understanding what is representative for a specific task is an iterative process. ML practitioners need to change data, models, and their associated processes as they become more familiar with their modeling task, as the state of the world evolves, and as products are updated or maintained.
Iteration directed by this evolving understanding seeks to improve model performance, often editing datasets to ensure desired outcomes.

Failure to effectively recognize data quality and coverage needs can lead to biased ML models~\cite{mitchell2020diversity}.
Such failures are responsible for the perpetuation of systemic power and access differentials and the deployment of inaccessible or defective product experiences.
Yet building representative datasets is an arduous, historically difficult undertaking~\cite{tayi1998examining, 49953} that relies on the efficacy of human-specified data requirements.

To ensure a dataset covers all, or as many, characteristics as possible, specifications must be the result of a comprehensive enumeration of possible dimensions---an open and hard problem that few have practically grappled with in the context of ML. Further contributing to this difficulty is the realization that it is not enough for the training datasets to be aligned with expected distributions: they must also include enough examples from conceptually harder or less common categories if said categories are to be learned~\cite{asudeh2019assessing}. Failure to sufficiently consider both the critical dimensions of data and their relative complexity can have troubling consequences. Instances of such missteps span issues of justice, healthcare, hiring practices, voice and face recognition, and loan qualifications, wherein biases of data and algorithms limit technological use and cause harm ~\cite{buolamwini2018gender,asudeh2019assessing,palanica2019you,angwin2016machine,noble2018algorithms}. Yet understanding these data requirements even after training is difficult; knowing them \textit{a priori} is exceptionally so.

Rather than emphasize tools that enable better collection and data iteration practices---that \emph{design better data}---research in fairness and machine learning has largely focused on prescriptive ``how-to'' frameworks,  definitions of fairness, and post-collection analysis techniques~\cite{amershi2019guidelines, yang2020towards}. While there are exceptions to this~\cite{hohman2020understanding}, the hidden technical debt~\cite{sculley2015hidden} accumulated from poor data design remains an under explored space. To reduce this technical debt and encourage diverse datasets, methods of externalizing 
data collection, iteration, and training are necessary checks for ensuring datasets reflect diverse experiences and are robust when deployed in real-world models. 

\textbf{Contributions}   We present \emph{designing data}, an iterative, bias mitigating approach to data planning, collection, and ML development; we implement this approach in an interactive dashboard that scaffolds each step for practitioner use. 
Motivated by 24 formative interviews \ref{sec:informal}, designing data is a structured, holistic parallel to the current standards for developing production-quality datasets~\cite{hohman2020understanding}. Each step proactively introduces interventions to improve models prior to deployment: 
\begin{enumerate}
\vspace{-2mm}
    \item \emph{\textbf{Pre-Collection Planning}} prioritizes reflexive consideration for domain and data needs prior to modeling, documents expected distributions of attributes, and highlights potential biases through questions related to class or characteristic representation within data. 
    \item \emph{\textbf{Collection Monitoring}} communicates insight into dataset evolution, allowing users to make targeted adjustments to collection processes based on new insight or disparities between expected distributions and existing data. 
    \item \emph{\textbf{Data Familiarity}} borrows from Out-of-Distribution (OOD) methodologies to identify data that a model perceives as unfamiliar, creating new directives for data and model iteration. 
\end{enumerate}

We demonstrate designing data's effectiveness through a case study using inertial measurement units (IMU) data--time series data representing X, Y, \& Z positioning--to classify hand position while texting. First, we collected data iteratively, using Pre-Collection Planning and Monitoring steps to build a diverse dataset. Then, we use leave-one-out cross-validation to evaluate how these initial steps  influence performance. Finally, we evaluate familiarity to first debug a dataset, then to direct collection efforts. Each step is centralized within a dashboard. We find models trained on highly diverse data outperform those trained on less diverse data across intersectional groups.

\vspace{-.2cm}
\section{Related Work}
\paragraph{ML Documentation}
A close alternative to \textit{designing data} is model and dataset documentation, such as Model Cards \cite{mitchell2019model} or Datasheets for Datasets \cite{gebru2021datasheets}.
Such work details what to include within said documentation for transparency in downstream model and dataset use \cite{Koesten2019, bender2019data, arnold2019factsheets}.
This type of documentation has been widely adopted \cite{hopkins2021machine}. 
However, documentation on its own is limited, as it typically accompanies a model or dataset upon release rather than shaping its development.
For instance, Model Cards do not explicitly guide how to reconcile model weakness or actively direct data collection to improve fairness--rather, they are intended for transparency.
Our contribution is embedding the transparency revealed by such documentation into our designing data process and dashboard. Through simple prompts, users actively surface and engage with their priors--which later informs their evaluation.

\textbf{Reflexivity and Self-Reflection}
In social science, the practice of \textit{reflexivity} is a way
to externalize implicit subjectivity present in data collection and interpretation \cite{fish2021reflexive, dodgson2019reflexivity}. Reflexivity entails deliberately examining practitioners' own assumptions, practices, and belief systems, then \textit{contrasting} them with alternative perspectives. This acknowledges positionality---how differences in social position and power shape identities. While reflexivity is typically practiced retrospectively, 
~\cite{soedirgo2020toward} outlined how it could be an active process through \textit{``ongoing reflection about our own social location and [...] our assumptions regarding others’ perceptions.''} This approach includes recording assumptions of positionality; routinizing reflexivity; including other actors in the process; and communicating reflexive outcomes with data. 

Separate from yet related to reflexivity is recent data visualization work prompting viewers to reflect on their individual beliefs of data
\cite{8019830, hohman2020communicating}. Examples of this include The New York Times ``You Draw It'' visualizations, where readers draw a projected trendline of what they think the data looks like and then compare that projection with the real data \cite{aisch2015you, katz2017you, buchanan2017you}, and an MIT Tech Review article illustrating the complexity of building fair recidivism models wherein readers change different hyperparameters then contrast their outcomes to existing models \cite{hao2019can,strayinteractive}. Such visualizations act as powerful tools ensuring self-reflection---an important factor in building representative datasets.

\textbf{Learnability and Familiarity}
Having diverse characteristics with sample size parity in data is one step towards mitigating bias but does not ensure equitable learning across classes. Instead, these approaches ignore data learnability: having an equal number of samples per class neglects the fact that some classes are inherently easier to learn than others~\cite{ben2019learnability, schapire1990strength, klawonn2019exploiting}.
Unfortunately, this complexity might not become apparent until after deployment.
In order to build better systems, discovering what has not been learned by a model is critical. In the past, testing data acted as a proxy for this evaluation. But while test datasets offer excellent thresholds for performance expectations, these are not the same measurements: testing accuracy measures whether a model was correct in its classification without consideration for how it reached its conclusion. For this reason, models are often poorly calibrated \cite{van2019calibration}. 
To address this, we use density estimation (DE)--a common technique for measuring OOD--to evaluate how familiar a model is to a given sample. Our implementation of DE is most similar to \cite{lee2018simple}, who use it to detect anomalous samples \textit{post}-training (e.g., adversarial attacks). 
Unlike prior work, we extend DE techniques into training to direct data efforts.

Our use of DE to guide data collection is a response to two drawbacks we note from other work seeking to improve data representation and model performance: (1) alternative approaches require \textit{prior awareness} of what facets of data are variable---for data that people are unaccustomed to, this might be impossible---and (2) measuring diversity within a given dataset does not account for what the model actually \textit{learns} \cite{hooker2021moving, schmidt2018adversarially}.
Because it is \textit{model} outputs that we are concerned for, these are important weaknesses to counter.

\vspace{-.1cm}
\section{Formative Interviews}
\label{sec:informal}
To understand ML practitioners' data collection needs, we conducted 24 semi-structured exploratory interviews with individuals possessing extensive machine learning and data collection experience within a large technology company. The interviewees ranged from ML research scientists and designers focused on ML experiences to engineering product managers. Interviews lasted $\sim$ one hour. 
As interviews progressed, common themes surfaced that directed our attention to issues of data coverage and representation. We describe those themes (T1-T3) below.

\textbf{Critical dimensions of data are hard to know a priori (T1).}
A proactive approach to data collection requires knowing what axes are important for observation. As one interviewee put it, \textit{``how do we know who and what is missing?''}  This was a shared difficulty described in nearly all interviews. While it is typically impossible to have a complete understanding of critical dimensions of the data before starting data collection, there are some common characteristics for human-centric data collection based on existing knowledge of population statistics and power imbalances. How useful that information is depends on the context of how the data is used---for example, a person's accent and speech pathology are important factors in speech recognition, but not for creating a personalized wine recommendation. Unlike summary statistics, surfacing noisy data or missing subsets--essentially ``debugging'' data-- requires significant effort. As another interviewee said, \textit{``fairness analysis is useful to a point''}. Generic tools for surfacing these nuanced limitations of data do not exist, but there is both need and desire for them~\cite{holstein2019improving}. 

\textbf{Difficulty of collection leads to compromises (T2).}
Data collection is a difficult process to launch, requiring significant tooling. This difficulty contributes to issues of representation in data, as the emphasis in early data collection is on \emph{how} to collect and structure data rather than building a complete picture of \emph{what} to collect for. Further, early collection efforts tend to prototype, making convenience sampling canon. In human-centric data (e.g. speech or movement), such requirements can encumber diversity across many axes. This has the potential to inhibit robust ML---while there is a natural iteration in datasets stemming from distribution shift~\cite{quionero2009dataset} and collectors' evolving understanding, it is a cycle full of forking paths~\cite{kale2019decision, hohman2020understanding} and dead-ends.
 
\textbf{{Model failures are invisible without participation and iteration (T3).}}
Real world failures are only visible when communicated. But that communication comes from those invested enough in the tool, system, or research agenda to make the effort to bridge the communication gap. To quote an interviewee: \textit{``The people who had issues were invisible to the system because they didn't like using it''}. Unlike other domains which employ tools to surface and track issues via user engagement, there are currently no tools that address the gap between a deployed ML product (let alone its early prototype) and a user. Such gaps are widened by language and knowledge barriers, and the products themselves are inaccessible to many who do not align with the priors on which the system is built. As these limitations have substantial downstream impacts, we sought to introduce early, comprehensive data \& model checks to facilitate easy pivoting during data collection.

\begin{figure*}
\vspace{-.2cm}
  \centering
\includegraphics[width=.7\linewidth]{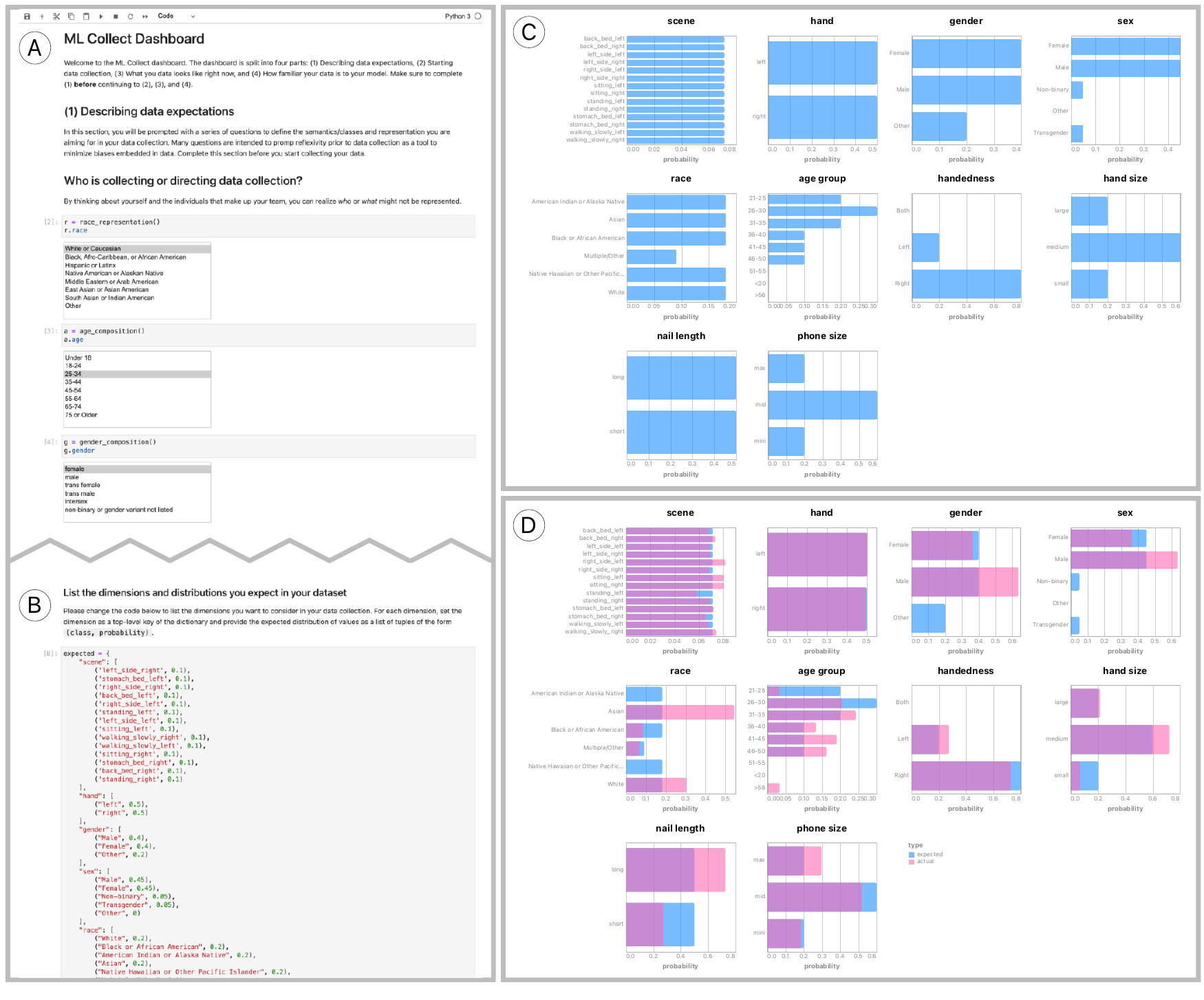}
  \caption{
  (A) Dashboard prompts practitioners through designing their datasets.
  (B) Before collection, the dashboard prompts documentation of expected distribution.
  (C) These are visualized as histograms.
  (D) As data is collected, true distributions (pink) overlay expected distributions (blue), highlighting divergent patterns.
  \vspace{-.6cm}
  }
  ~\label{fig:expected}
\end{figure*}
\vspace{-.2cm}
\section{Designing Data}
\label{sec:designing}
Existing bias mitigation and model evaluation approaches attempt to address the themes uncovered in our formative interviews but are not comprehensive to data collection \emph{and} machine learning pipelines.
In response to this disparity, we propose an iterative approach to data collection and machine learning that we call \emph{designing data}. Designing data responds to our themes by introducing interventions before, during, and after collection \& training. Shown in~\autoref{fig:designing_data}, each step is intended to complement the others, compensating for their limitations to holistically improve the development and deployment process. 
We describe each below.

\textbf{I: Pre-Collection Planning}
To facilitate broader consideration of critical facets of data, data design requires explicit documentation of \emph{what} will be collected---including expected dimensions and distributions of data---\emph{before} collection begins. Our documentation process ensures developers pay close attention to data diversity and coverage early in an ML pipeline and creates reference points for comparison when new information is uncovered.
This process aligns with prior work in heuristics, implicit bias, cognition, and fake news susceptibility: deliberation---\emph{reflection}---can correct intuitive mistakes, such as those made in data collection~\cite{pennycook2019lazy, bago2020fake, toplak2011cognitive, saul2013scepticism}. While it is difficult to enumerate all critical factors, this first step in our designing data process provides scaffolding, responding to our first theme (T1): \textit{critical dimensions of data are hard to know a priori}.

As shown in~\autoref{fig:expected} (A), Pre-Collection Planning asks a series of questions to prompt reflexive consideration for individuals' personal biases through a series of open-text prompts, drop-down selections, and simple declarations. When the data relates to a human subject (\ie images of faces, movement data, or voice recordings), our dashboard prompts self-reported demographic information about teams or individual users involved in developing the dataset. 
After, users are asked what data distributions they expect. This is an important step: prior work suggests that simply recognizing such information~\cite{fish2021reflexive, kim2017explaining}, and introducing design frictions~\cite{gullstrom2012design, pennycook2020fighting}, may improve the quality of data work. An example of the dimensions and related distributions is shown in \autoref{fig:expected}. From these inputted dimensions, audits of population statistics, missing data, and undersampled subsets are presented to users. 
Different categories of the data, including demographics, metadata, task specifics, class representation, and intersectional categories are visualized.

\textbf{II: Collection Monitoring} By stating expected distributions prior to collection, auditing the data against those distributions is straightforward, allowing readjustments to be made quickly when necessary. Each view reflects the data's evolution, allowing real-time insight on where additional collection is needed. Our dashboard includes graphs highlighting distribution disparities that were perpetuated or introduced as the data was collected, as shown in ~\autoref{fig:expected} (D). These charts benefit users by noting when the data collection process is either skewed (for example, through convenience sampling) or when previously stated expectations did not align with reality. This step's iterative nature shortens the response time to correct fundamental errors in data, highlighting limitations that may have remained unnoticed, as described in our second and third themes: \textit{difficulty of collection leads to compromises (T2)}, and \textit{model failures are invisible without participation and iteration (T3).} 

\textbf{III: Data Familiarity}
 After a model is trained, understanding what it has and has not learned appropriately is critical \footnote{The purpose of designing data is not to replace existing tools but rather to encourage a holistic approach that incorporates new ML techniques. For interventions \textit{during training}, we refer to existing literature~\cite{murphy2012machine,japkowicz2002class}.}. The value in this form of auditing is significant: some data may prove more difficult to learn due to their inherent complexity, or from not having enough similar examples. Despite the increased rigor of data collection as result of our earlier designing data steps, expected and actual data distributions might not match the learning needs for the model, thus requiring a stop gap such as under or oversampling, generating synthetic data, or continued data collection. We adopt density estimation (DE) to measure how familiar our model is with individual data points. While DE for OOD detection is well studied \cite{gawlikowski2021survey}, our use of DE to direct data work (e.g. collection and annotation) is unique. 
By gaining insight on how a partially trained model perceives data, we can focus efforts on the most useful subsets, reweighting or replacing the data accordingly. In this way, even though \textit{difficulty of collection leads to compromises (T2)}, we are able to interactively uncover remaining issues, as suggested by (\textit{T3}).

Unfamiliar samples are ``edge cases''---those that either are not represented appropriately within the dataset, are particularly challenging for the model to learn, or were erroneously collected (e.g., noisy). In early dataset development, familiarity scores act as useful checks for data with little signal---samples where we expect the model to perform well yet \textit{may} be noisy and thus require human inspection . To measure the familiarity of the data, we incorporate density estimates of layer activations across a neural network (NN). We focus on the penultimate layer before the prediction softmax as it is the final feature representation used to make a prediction. Passing $N$ inputs through the network produces an activation matrix $A(N) \in \mathbb{R}^{N\times M}$ for all $M$ neurons in subset of selected layers $L'\subseteq L$. We learn a Gaussian Mixture model on these layer activations as given by the following:
\vspace{-.1cm}
\begin{equation}
  p(x|\lambda) =  \sum_{i=1}^M w_i g(x \mid \mu_i, \Sigma_i)
\end{equation} where $x$ is a matrix of layer activations, $w_i, i = 1, . . . , M$ are the mixture weights, and $g(x|\mu_i,\Sigma_i),i = 1,...,M$ are the component Gaussian densities \footnote{This mode of DE is interchangeable with other DE or OOD methods such as \cite{wkeglarczyk2018kernel, lee2018simple}.}.
For each sample in the current training set, we obtain the activations from layer $l$, then use PCA to reduce its dimensionality. This projection step serves two purposes: it reduces the dispersion of points that is typical in high dimensional spaces 
, and makes the remaining computation more tractable.

We then perform a Variational Bayesian estimation of a Gaussian mixture~\cite{zobay2014GMM} in the projected space. The fitted GMM allows us to give a familiarity score to each new sample. Given this sample, we  extract the activation from the same layer $l$, again apply dimensionality reduction, then evaluate the log-likelihood provided by the fitted GMM---this is our familiarity score. If the sample falls into a densely populated area, its log-likelihood will be high: from the perspective of the features extracted by the current state of layer $l$, this sample appears as \textit{familiar}. Conversely, if the sample falls into a scarcely populated area, its log-likelihood will be low and the sample less familiar. This measurement can be applied to new samples---those not seen by a model previously---or on training samples. Familiarity scores are presented in our dashboard through a series of graphs depicting their range and frequency, providing users direction for future data efforts; samples that are unfamiliar are reconciled with human expectations.

\vspace{-.2cm}
\section{{Task Selection, Data Collection, Modeling}}
\label{sec:methods}
We instantiated our designing data approach through a human activity recognition (HAR) task using \emph{inertial measurement unit} (IMU) data---
While designing data generally applies to all data collection and machine learning processes, our selection of data type and task were motivated by the unique challenges IMU data presents to building data diversity. First, IMU data inherently lacks the closeness of mapping~\cite{blackwell2001cognitive} that image and audio data have to human models of the world, making it more difficult to audit. Recognizing when IMU data coverage is incomplete can be difficult when compared to image data as levels of abstraction often obfuscate fundamental problems in a dataset~\cite{ramasamy2018recent, bartram2021untidy}). Second, IMU data requires contextualization to create meaning---real-time labeling, or additional information from audio and video---and is harder to collect, unlike images and audio clips which are now ubiquitous~\cite{ramasamy2018recent}. Yet IMU data still has the potential to bias ML models. Thus, we evaluate designing data's merit within a typically challenging context.

Our experimental task resembles work by~\citet{goel2013contexttype}, who improved mobile text entry by categorizing different hand positions when users typed. 
Before collection, participants provided demographic information and metadata. This included race, ethnicity, gender, sex, age, hand length (mm), nail length (mm), hand dominance, phone version and phone size. 
We collected data from 33 participants recruited over three separate periods in response to data disparities highlighted by our dashboard. 
In total, we collected $>3.88$ million measurements from $1455$ sessions.
This data is used to populate dashboard, train and evaluate classifiers, then refined according to familiarity evaluations. We use 
1D convolutional neural nets (CNNs) for classification \footnote{See appendix for details on data processing, modeling, and iterating with Pre-Collection Planning \& Monitoring}.

\vspace{-.2cm}
\section{Modeling Experiments}
\label{sec:modeling_experiments}
The ability to produce diverse subsets does not guarantee appropriate representation of sensitive attributes. 
Our approach to encouraging fair outcomes considers the amalgamation of both data and model. The following experiments adapt the following definition of fairness---rather than seek a high \textit{average} accuracy across classes, we look at nuanced performance---accuracy, loss, and misclassification---between intersectional groups. To this end, we consider several questions as part of our designing data evaluation:  (\textit{Q1) Does auditing  to increase data diversity improve model generalizability?} and \textit{(Q2) Is data familiarity useful in auditing model \& data?} 
We evaluate our interventions through a series of modeling experiments. First, we show that diverse data \textit{does} lead to better performance. Then, we use familiarity to uncover noisy data within the dataset and describe how removing these samples impacts intersectional accuracy. Finally, we show that supplementing the dataset with unfamiliar samples improves model performance.

\vspace{-.2cm}
\subsection{Diverse Data: Does auditing to increase data diversity improve model generalizability?}
\label{sec:diverse-data}
We sought to answer Q1 through our first set of experiments. We compare ``diverse'' models to ``less diverse'' models. We do so for two reasons: first, it may not be clear to practitioners that collecting diverse data early in development is critical to building functional tools. Second, despite best efforts to curate a list of meaningful characteristics, we did not know if these additional data dimensions had any true effect on the classification task.

Both diverse and less diverse models are trained using the same number of training samples and are evaluated on the same test data. Less diverse models are trained on data where one group (e.g. small handed) was left out. In this way, we perform  leave-one-out cross-validation \textit{and} consider the specific effects of a given demographic group. All models were trained with the same sample size.
We then generate predictions from the original test set.  Paying close attention to intersectional groups, we expect to see more performance stratification in less diverse models.
We compare models across overall, group-specific, and intersectional accuracies. We hypothesize that some categories are more meaningful to performance than others. Performance disparity---such as lower accuracy---across categories \textit{despite} equal sample size would support this hypothesis.

\begin{figure*}
\vspace{-.8cm}
  \centering  \includegraphics[width=.9\linewidth]{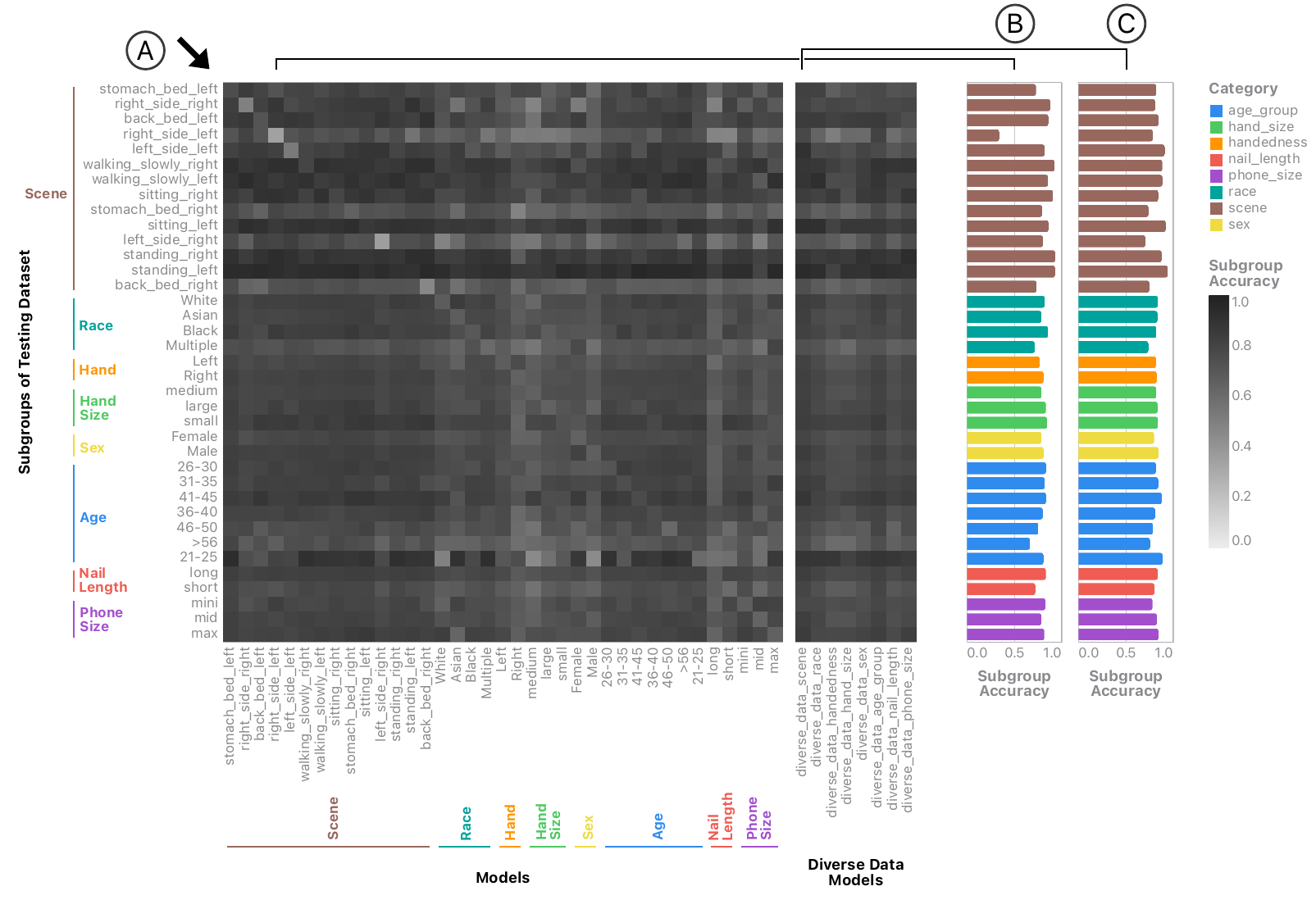}
  \caption{
    Performance per subgroup on testing data (y-axis) per model (x-axis), split across categories.
    The large square matrix (left) shows accuracy  where a subgroup was left out of the training set. The rectangular matrix (right) shows performance of models trained on diverse data.
    (A) The arrow highlights the diagonal of the matrix: subgroups of data that perform worse than others, corresponding to the model trained without this particular subgroup.
    (B) E.g., taking the \texttt{left\_side\_right} model from the matrix,
    we see it performs poorly on the \texttt{left\_side\_right} subgroup.
    (C) Models trained on same sized yet more diverse dataset show no dip in accuracy. 
    \vspace{-.7cm}
  }
  ~\label{fig:all-1d-subgroups}
\end{figure*}

\begin{figure*}
  \centering
\vspace{-.5cm}  \includegraphics[width=.9\linewidth]{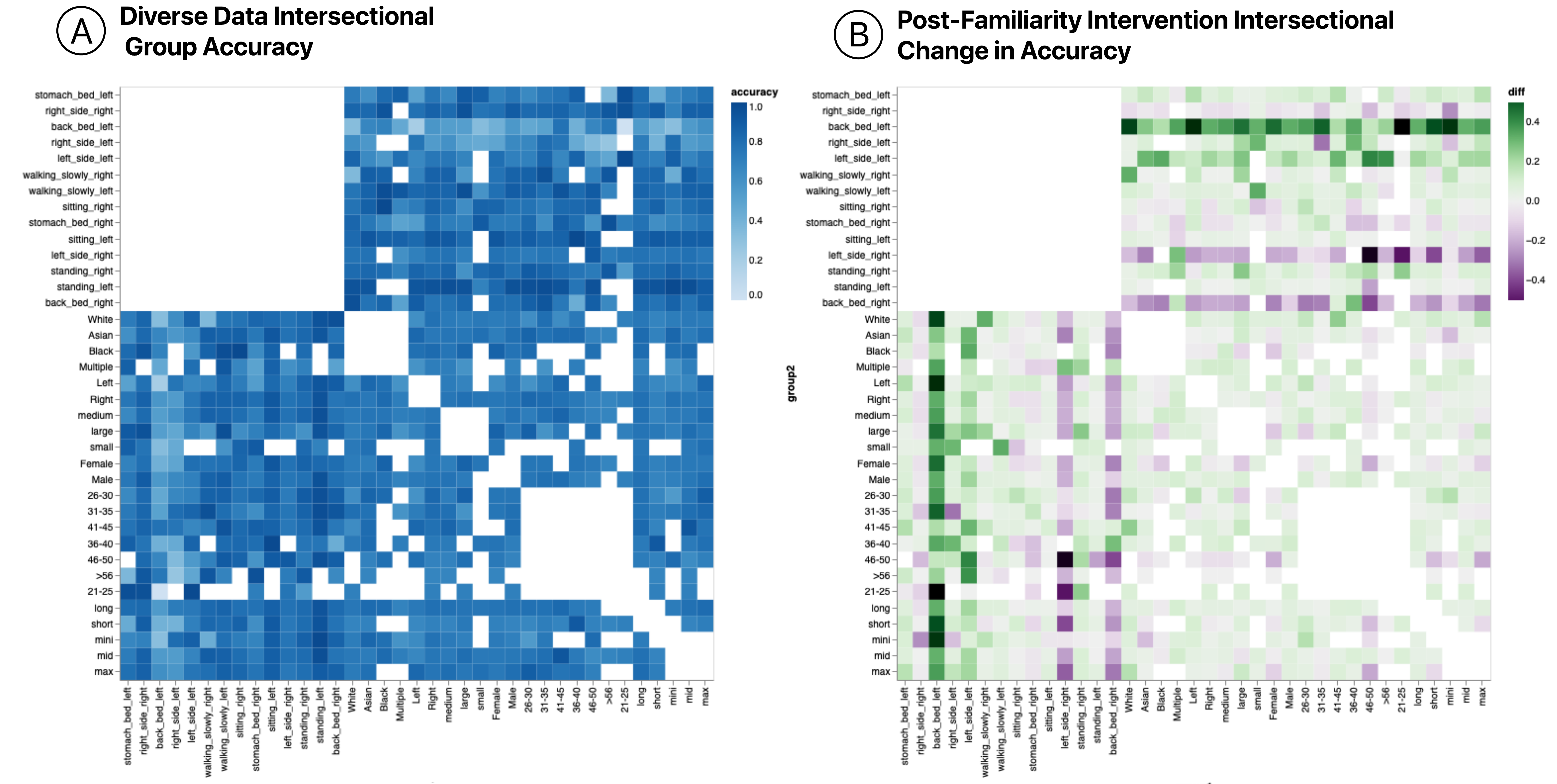}
  \caption{
  (A) Accuracy for intersectional subgroups in a diverse data model.
  (B) Difference in accuracy for subgroups before \& after familiarity interventions (green is positive, purple negative).
  \vspace{-.9cm}
}~\label{fig:familiarity_change}
\end{figure*}

\textbf{Evaluation} When compared to models trained on less diverse data, we found that models with diverse data had an overall higher accuracy \textit{and} performed better across intersectional groups, as shown in ~\autoref{fig:all-1d-subgroups}. For example, ~\autoref{fig:all-1d-subgroups} B shows a significant dip in performance for the model trained without data from the \textbf{right-side-left} condition compared to diverse model (\autoref{fig:all-1d-subgroups} C). This is as we anticipated, and we see that for data where participants were typing in the \textbf{right-side-left} condition, the less-diverse model actually performs  \textit{worse} than random chance. This pattern is repeatable---across $k$ models where we intentionally left out one group  (\autoref{fig:all-1d-subgroups} A), we see a correlated diagonal of lighter color indicating lower testing accuracy, supporting our hypothesis that the extensive characteristics we collected data for \textit{do} effect IMU performance. In contrast, diverse models show less performance variance, instead performing better across different demographics---matching what we would hope to achieve to minimize worst group generalization \cite{sagawa2019distributionally}.
We found some intersectional subgroups performed drastically different compared to their overall group performance. In typical evaluations, this nuance is often obfuscated by aggregation, yet we were able to capture it using simple interventions and visualizations.

\vspace{-.1cm}
\subsection{Familiarity: Is data familiarity useful in auditing model \& data?}
\label{sec:familiarity}
There are two scenarios where familiarity is useful in directing dataset iteration: to facilitate data cleaning, and to encourage appropriate representation of diverse data for a given model. When a dataset contains noisy data, familiarity can help surface these samples for human evaluation. If a dataset has already been cleaned, then familiarity is used to highlight samples that less familiar. We evaluate familiarity's efficacy in directing dataset iteration (Q2) through a series of experiments to improve performance on the same testing data used in ~\autoref{sec:diverse_data_exp}, modifying training data but maintaining sample size parity. Modifications to the training data are completed using ``self-familiarity'' scores (i.e., familiarity of training data) to avoid overfitting. 
\vspace{-.3cm}
\paragraph{6.2.1 Familiarity for Debugging}
\label{sec:debugging}
Our dataset was collected ``in the wild'' without additional annotation from participants. While participants were given clear instructions, we anticipated that a small number of samples would show significant noise or distortion---participants might drop their phones as they type, or switch hands part way through a session. Such samples may introduce unwanted effects downstream if left uncaught. Given the size of the dataset, inspecting time series plots for each session was untenable. Instead, incorporating familiarity may \textit{greatly} reduce the number of samples to evaluate.
By using an least partially trained model on the dataset to uncover unfamiliar samples, familiarity offers a possible solution to capture noisy data,
presenting an alternative to work such as ~\cite{pleiss2020identifying, shen2019learning}, which incorporate loss as a metric for capturing noisy samples. We hypothesize that least familiar samples within a noisy dataset will include instances of noisy data  posing the greatest harm to the model.

We explored how to incorporate familiarity as a tool for debugging first through an automated approach to removing data, then by incorporating human review. Our protocol is as follows: first, we train an initial model on all available training data. We apply self-familiarity to the training and testing set, selecting only 0.1\% of the data corresponding to least familiar samples. 
This data is either removed from the dataset, or visualized and manually reviewed per sample to evaluate if they are truly noisy or simply uncommon. We removed the same number of samples with manual review of truly noisy data as with the automated removal \footnote{Experimental protocol can be found in the Appendix.}.
We compared outcomes of both automated removal and removal through human review, but found that deleting a percentage of least familiar data removed both noisy data and important outliers. We evaluate this outcome by comparing the results of the following experiment to improve data diversity across each approach to data cleaning when compared to a baseline where no data is removed.

\textbf{Evaluation}
In practice, we found that familiarity worked well as a tool for debugging. Before removing noisy data from the complete dataset, a large percentage of the least familiar samples showed significant distortion, despite our efforts to normalize the data. Because of the presence of these noisy data, running our familiarity experiments with data that was not cleaned did not show the same levels of general improvement. In this case, matching metadata to noisy samples was not the correct comparison---these samples were not exemplars of the subpopulation. 

In manually evaluating our unfamiliar data before removal, we uncovered cases where unfamiliar samples were not noisy but rather underrepresented intersectional groups. For instance, a person identifying as an \textbf{Asian} \textbf{Female} with \textbf{Small} hands and \textbf{Large} phone was as unfamiliar to the model as incredibly noisy examples.  Noisy data has different implications for the model than  unusual samples. For this reason, removing 0.1\% of the least familiar data prior to experiments did not improve performance to the extent seen when manual review was implemented. That is, familiarity cannot distinguish between distorted noise and underrepresented or OOD data. For datasets where significant noise is present, \textit{human review} is necessary to evaluate data quality.

\vspace{.2cm}
\textbf{6.2.2 Familiarity for Diverse Data Coverage}
\label{sec:diverse_data_exp}
Next, we compare familiarity scores across the different descriptive groups, as described in ~\Cref{sec:diverse-data}. These scores are used determine next steps for additional data collection, augmentation, or modified (over/under) sampling. Here, we sample \textit{out} a percentage of the \textit{most} familiar data, and \textit{add} data matching the intersectional characteristics for the same percentage of \textit{least} familiar data. Substituted data was held out from training in ~\Cref{sec:diverse-data}, thus is new to the models. We replace samples based on metadata characteristics, using combinatorial optimization to best match unfamiliar data.
Using PCA, we project down to 50 dimensions, then fit 5 GMMs to last dense layer for each 1D CNN trained in ~\Cref{sec:diverse-data}.
Scoring our training data, we save familiarity scores and model weights. We vary the range of familiarity scores to sample from, percentage of data, and two sampling in methods---top $k$ and random selection from a least familiar data range---compared to a random baseline. We structure these experiments---varying window size and sampling percentage---to uncover a sweet spot: removing too much data may harm performance on familiar groups, and too large a window might impinge on less familiar data. 

\textbf{Evaluation}
Self-familiarity scores create a distinct curve, with unfamiliar data falling into the long-tail.
We find accuracy scores are far more striated prior to familiarity interventions, showing some concepts are learned better than others. Following familiarity interventions, models do show targeted improvement. Models performed more poorly in regions with high numbers of low familiarity data (an example of which is shown in ~\autoref{fig:familiarity_change}). Of note, models did not necessarily improve overall---although this was frequently the case--- instead showing improvement in areas of low performance and regression in those with high performance. In ~\autoref{fig:familiarity_change}, we show model performances on intersectional groups: (A) is a model trained on the \textbf{diverse-data-scene} dataset with no familiarity intervention. We can see lower accuracies in \textbf{back-bed-left} and \textbf{right-side-left} compared to other subgroups. In contrast, \textbf{sitting-left} and \textbf{back-bed-right} had much higher accuracy compared to other subgroups. \autoref{fig:familiarity_change} (B) shows the difference between this model and one trained on the same data with familiarity interventions. Regions with relative poor performance improved dramatically, while those with higher accuracies showed some regression.  Incorporating 0.1\% least familiar data was optimal for our experiments. 
Overall, familiarity consistently captured under-represented samples.

\vspace{-.2cm}
\section{Conclusion}
We need processes that integrate data \textit{and} models in systematic, transparent ways. 
While each step of \textit{designing data} can be incorporated in isolation, the interventions are complimentary, compensating for the cascading effects of upstream and downstream missteps. 
Data is shaped by the perspective of the observer; our work highlights how systematic processes may curtail bias early in development.

\section*{Acknowledgements}
We thank our colleagues at Apple for their time and effort participating in our research.
We especially thank Donghao Ren and Halden Lin for their help building the iOS application and Kayur Patel for his guidance. Aspen Hopkins was partially supported by the Siebel Fellowship.

\bibliographystyle{icml2023}
\bibliography{citations}

\appendix
\section{Data Collection}
\begin{figure*}
  \centering
  \includegraphics[width=.80\linewidth]{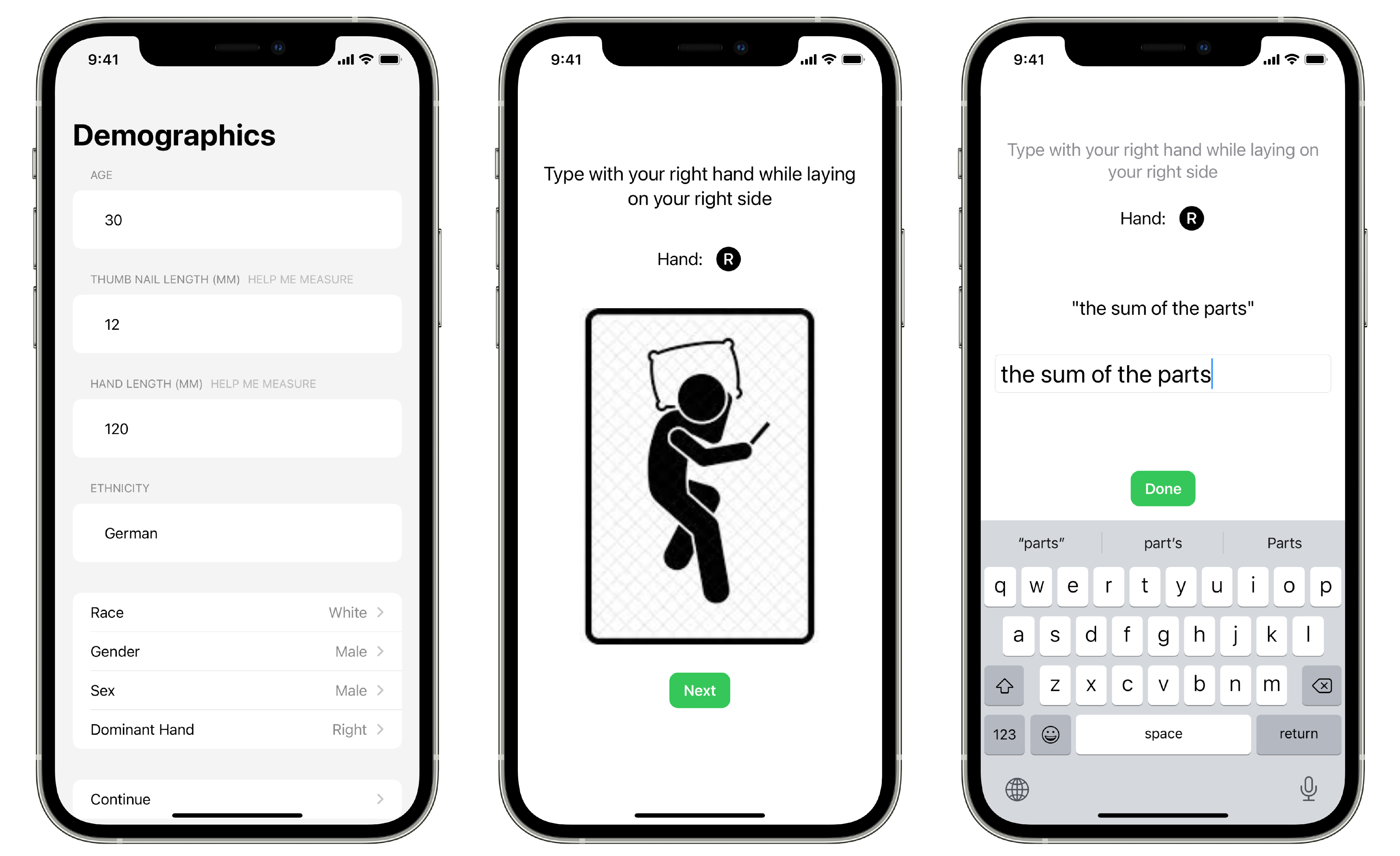}
  \caption{
  Our IMU Data Collection App showing the primary screens of the study.
  Participants entered their demographic information such as hand dominance, hand and nail length, sex and gender, and age (left).
  The app instructs a participant to type on their phone using either their right or left hand in a physical configuration (middle).
  Participants then type the presented phrase (right).
  }~\label{fig:app}
\end{figure*}

We built a custom iOS data collection app using the Swift programming language with public frameworks including SwiftUI, CommonCrypto, and CoreMotion.
The app collects right and left-handed texting data across different contexts people may type in, such as texting while walking or laying down. Example views of the app are shown in~\autoref{fig:app}. All data were collected with informed consent.
During collection, participants were given instructed how to hand posture when holding the phone. 
While this approach produces consistent data for training purposes, this limited the inherent variability within the data. 
We simplified the task to binary handedness (typing with the left or right hand), but introduced a source of natural variability by prompting users to perform a series of actions in parallel while typing. This contrasts ~\cite{goel2013contexttype}, who collected data while users sat in a lab environment. 
We introduced 6 typing scenarios (walking, sitting, standing, lying on your back, lying on your side, and lying on your front) to represent a selection of possible contexts users might normally experience typing. Participants were then asked to type an English phrase from MacKenzie and Soukoreff’s phrase set~\cite{mackenzie2003phrase} using their left or right hand for a total of $6 \mbox{ (positions)} \times 4 \mbox{ (typing sessions)} \times 2 \mbox{ (hands)} = 72$ trials. Entered phrases were recorded as a trial. Participants were automatically redirected to the iOS \emph{Measure} app to determine hand and nail length.
The app disabled keyboard autocomplete and autocorrect. 
Beginning when people pressed ``start'', we collected data was collected at a sampling rate of 200 Hz. When the task was finished, metadata, IMU data, and trial metadata (including scenario, phrase, keyboard recording, and session time) were pushed to an encrypted database.

\section{Dashboard \& Questions}
We built our dashboard in JupyterLab using Altair~\cite{VanderPlas2018}, a declarative  visualization library for Python, and Ipywidgets~\cite{Ipywidgets}. As shown in Main Body Figure 2 (A), the dashboard asks a series of questions to prompt reflexive consideration for the team or individuals personal biases. Using drop-down selections, they are asked to describe their team's representational make-up---including race, accessibility needs, age, sex and gender identities.
After filling out this information, users are told: \textit{``The following groups and their subsequent intersections are not participating in your project development. To ensure an optimal result, take steps to consider how their experiences and views might differ from the currently represented ones.''} This notice is followed by a list of demographic information not identified by the users. This is an important step: prior work has shown that simply recognizing such information~\cite{fish2021reflexive, kim2017explaining}, and introducing design frictions~\cite{gullstrom2012design, pennycook2020fighting}, can increase the quality and consideration that goes into data work. 

One limitation to this procedural approach is how scoped our questions are---they are not all-encompassing, but intended to start the process of reflection and inquiry early on. As a final step in this reflexive process, and to minimize this limitation, a series of open-ended questions that take in free-form text asks, ``What's missing, in the context of your project?'' followed by some examples that expand on axes of diversity teams might need to consider. Users are then asked to enter expected dimensions and distributions of data before collecting data (Main Body Figure 2 B,C). When distributions have incorrect values (\ie do not add up to 100\%), the dashboard normalizes. This active expression of expected data encourages users to acknowledge and document the specific limitations of their data, setting a precedent of conscious decision making from the beginning. It creates a simple provenance for early assumptions and a baseline to evaluate against during data collection.

From these inputted dimensions, audits of population statistics, missing data, and undersampled subsets are presented to users, reflecting \emph{During Collection}. New dimensions can be added as needed. Different categories of the data, including demographic information and metadata (described in the participant subsection), task specifics, and class representation are then shown in visualizations~\autoref{fig:expected}. Similarly, intersectional categories (such as age \textit{and} hand size) are shown. This view reflects the data evolution as more data is collected, allowing real-time insight into what additional collection might be required. Following collection, the dashboard allows users to use a pre-trained model or train their own. Following training, saved states of a neural network and model architecture are loaded into our familiarity functions. Data is inputted to build out familiarity scores, the final step in our designing data process. 

\subsection{Visual Examination of Familiarity}

\begin{figure}
  \centering
    \includegraphics[width=1.0\linewidth]{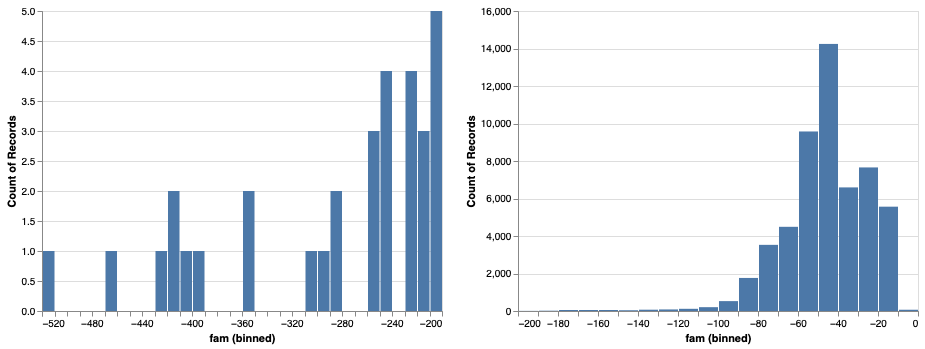}
  \caption{
  Overview of familiarity over the two tail ends of the distribution (most familiar on the right, least familiar on the left).
}~\label{fig:fam_overview}
\end{figure}

Log-likelihood is a relative measure. Given our interest in the spread of familiarity scores across different categories within the dataset, we introduced the visualizations shown in ~\autoref{fig:fam_before} and ~\autoref{fig:fam_before_2} into the dashboard for quick insight into if there are particular subsets that are seem less familiar than others. An overview of the familiarity scores can be seen in ~\autoref{fig:fam_overview} For example, with a close look we see that the \textit{least} familiar samples across genders are consistently \textit{female}. Similarly, we see our oldest age-range (>56) is least familiar for age. This view can be compared across iterations of data or model development for a gestalt of the changes in familiarity.

\begin{figure*}
  \centering
    \includegraphics[width=1.0\linewidth]{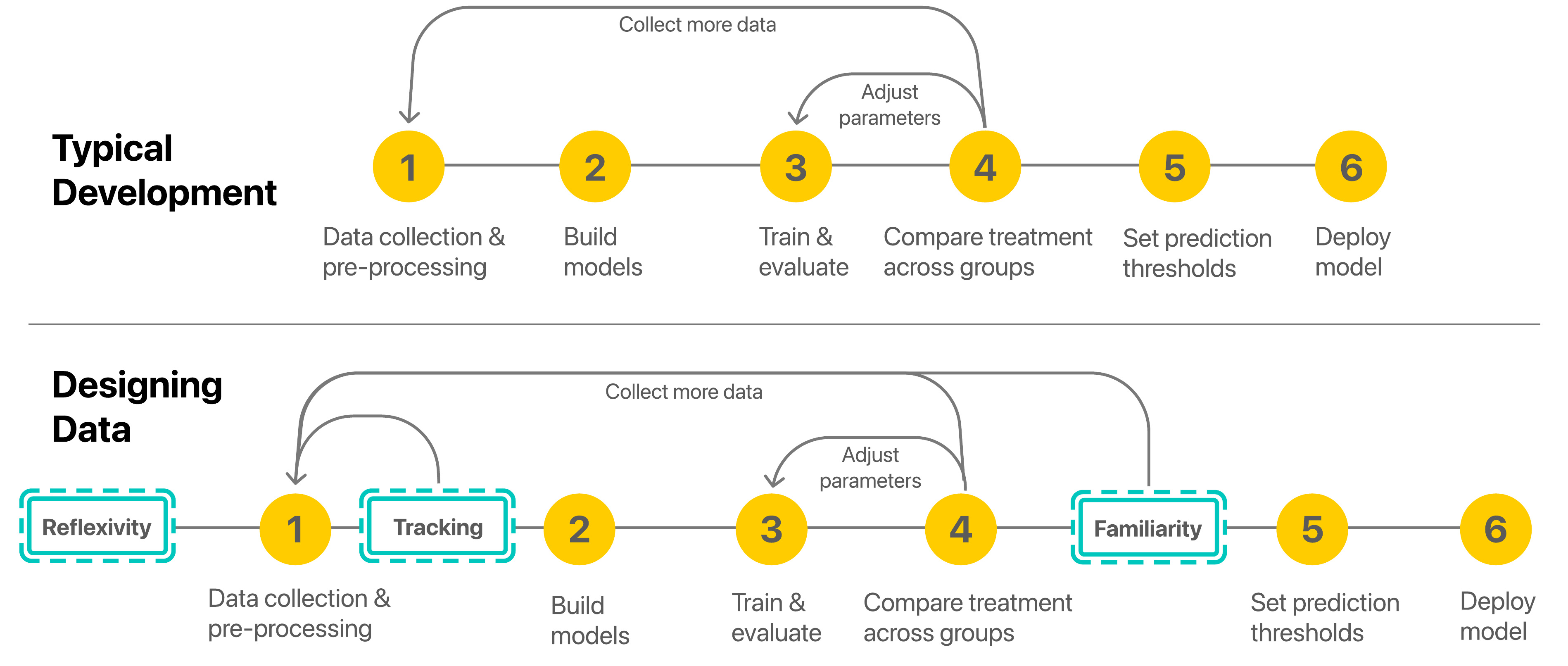}
 \caption{\textit{Designing data} process compared to conventional machine learning development. \textit{Reflexivity} ensures appropriate consideration of positionality and expectations is given before prior to collection. \textit{Monitoring} provides insight into unexpected trends during collection. \textit{Familiarity} facilitates debugging and highlights potentially noisy or underrepresented subpopulations to direct iteration. This figure represents a simplification of the data collection process. The results of \textit{Familiarity} can also be incorporated into training (after cleaning the dataset) and tracking (after training an initial model). 
 }~\label{fig:teaser}
\end{figure*}

\begin{figure}
  \centering
  \vspace{-1cm}
    \includegraphics[width=0.8\linewidth]{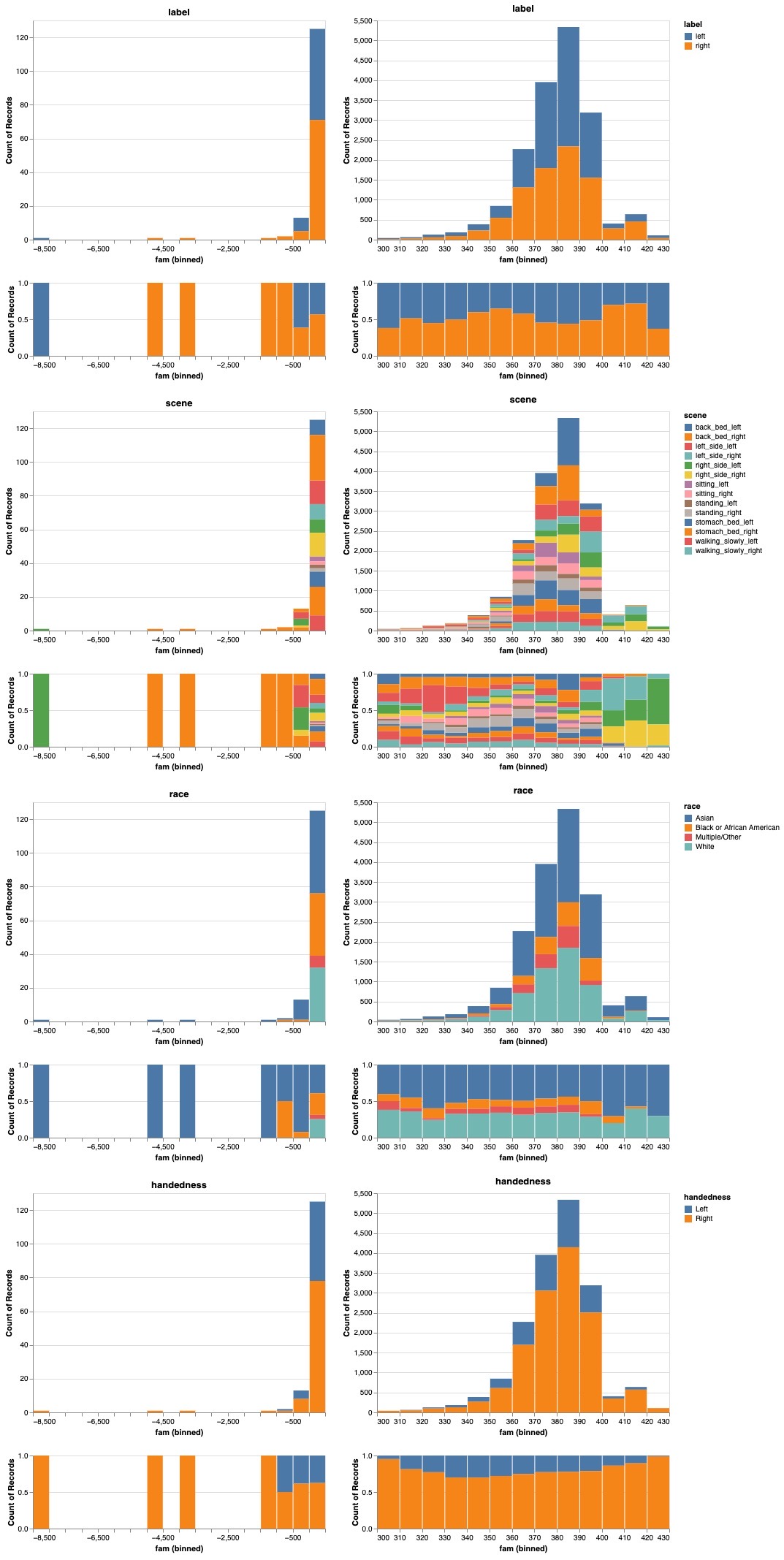}
  \caption{
  Visual presenting familiarity distributions across different categories (part 1).
}~\label{fig:fam_before}
\end{figure}

\begin{figure}
  \centering
  \vspace{-1.2cm}
    \includegraphics[width=0.68\linewidth]{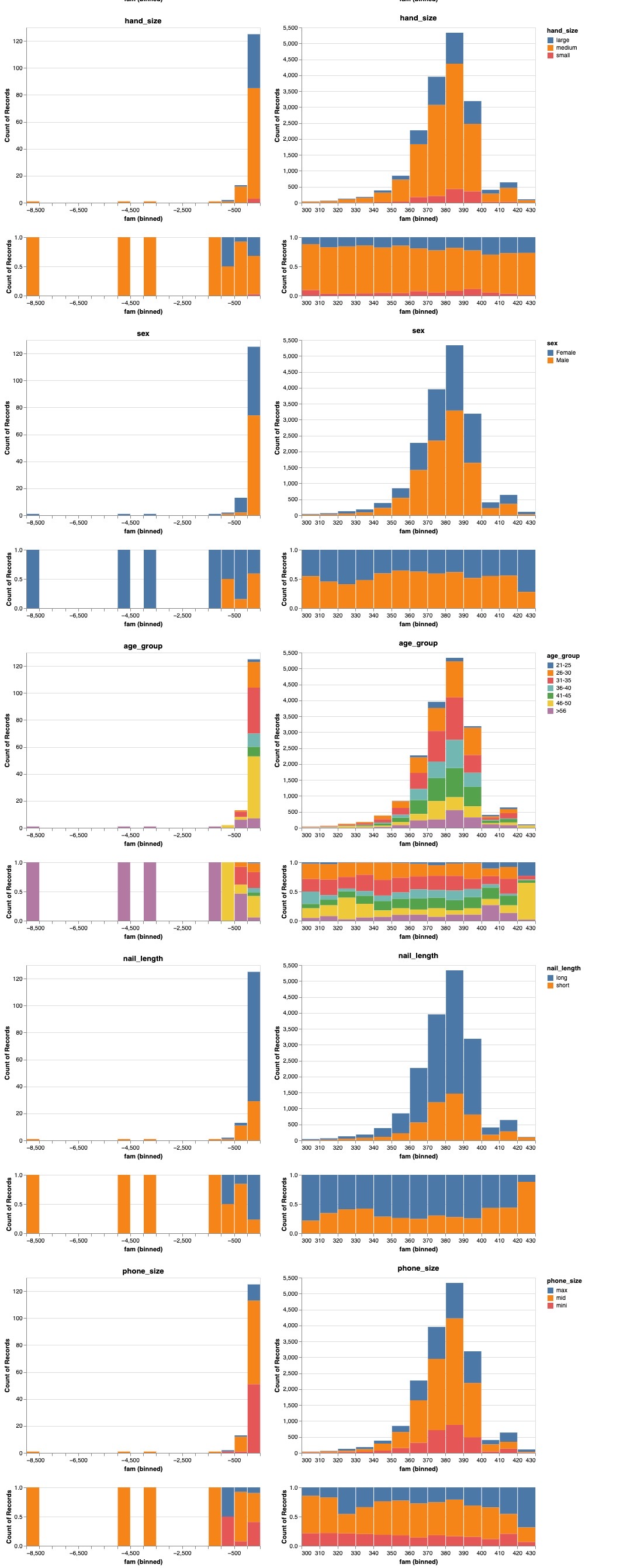}
  \caption{
  Visual presenting familiarity distributions across different categories (part 2).
}~\label{fig:fam_before_2}
\end{figure}

\section{Case Study: Reflecting on Our Data Collection Process}
\label{sec:case-study}

\begin{figure*}
  \centering
  \includegraphics[width=.9\linewidth]{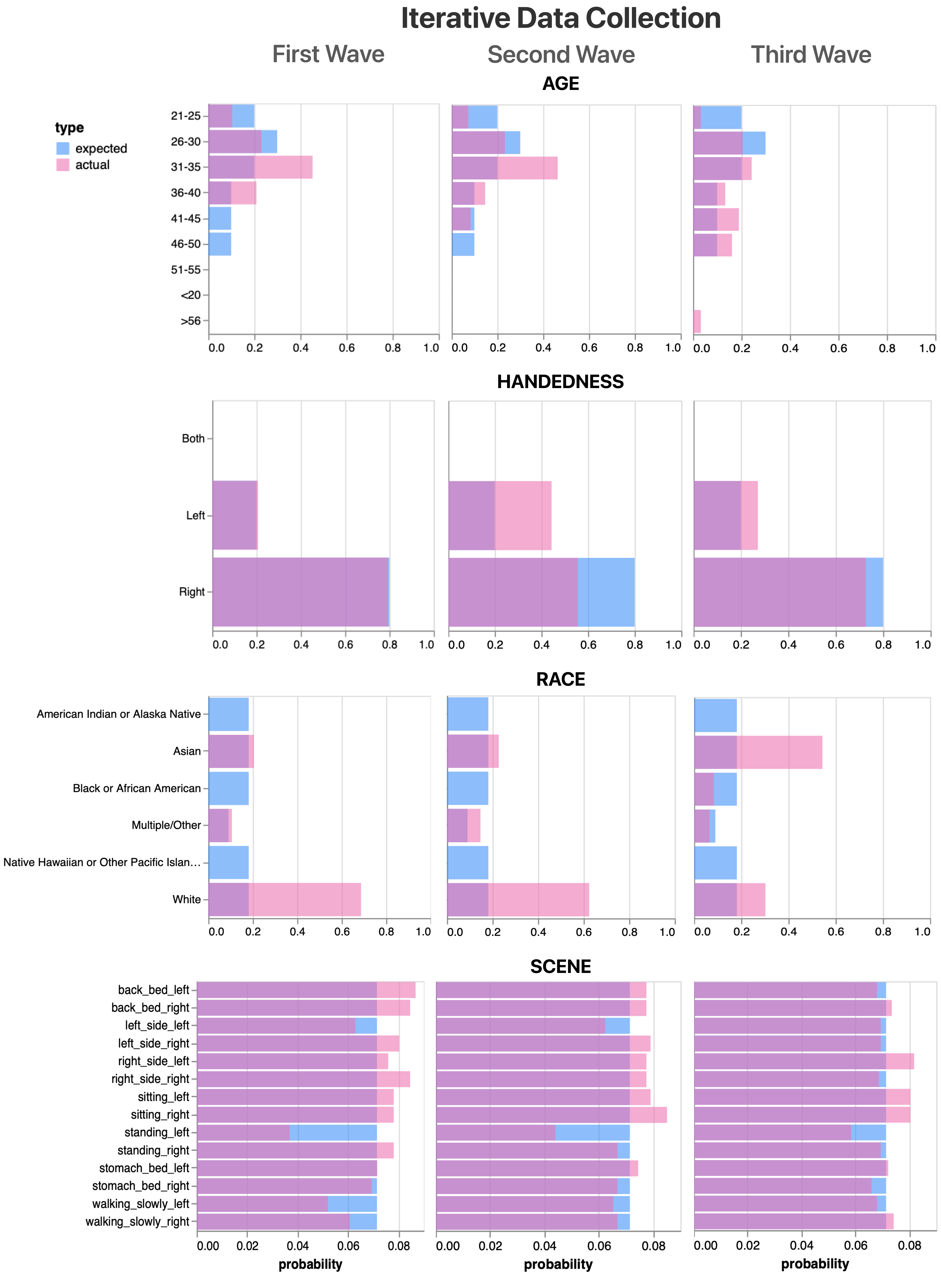}
  \caption{
    Selected views of data distributions across three different waves of data collection. Each new iteration of data collection was the result of evolving understanding about the data. For $Age$, initially there were no participants over the age of 38. $Handedness$ initially met expected distributions, however these did not match downstream needs. In $Race$, there was substantial skew for White participants. An error in data collection led to a skew in which scenes were presented to participants. 
  }
    \label{fig:collection_comparison}
\end{figure*}

It was unclear how diverse IMU data would influence our modeling experiments, or if the categories developed through our reflexive prompts would meaningfully align with variation within the data. Prior work on human gestures has shown age ~\cite{lee2020clip}, emotion ~\cite{noroozi2018survey}, and health \cite{8941232} influence gesture presentation.
The tilt of a smartphone during texting is captured by IMU data, and can be associated to back posture \cite{kim2020relationship}. Hand position during typing is similarly distinguishable \cite{goel2013contexttype}, yet IMU data collection rarely includes meta information about participants. Similarly, context is often not documented for image data (e.g. the proverbial question of \textit{what's outside the frame?}).

We explored how these and other common demographic categories influenced model performance and show how consideration for diverse data should be emphasized \textit{prior to} deployment, not just after deployment.
We incorporated our dashboard's \textit{Pre-Collection} suite of prompts in our own IMU data collection to determine what characteristics to collect for. It was through these prompts that we realized a need to measure additional information as typical demographics did not capture how people held their phones. We noted that \texttt{Phone Size}, \texttt{Hand Size}, \texttt{Handedness}, and \texttt{Nail Length} (particularly in the case of acrylic nails) may play a role in how people text, despite not being variables typically considered in such tasks. It was also through this process of reflecting on impactful features that we realized the advantages of asking participants to act out various behaviors during their typing tasks. This meaningfully shaped our task. Other measurements that were noted during this process would have added substantial complexity to our collection procedure. Most prominent of these was hand strength and dexterity. These features are impactful to how individuals type---a person with carpel tunnel or arthritis will type differently compared to someone without these conditions---but required additional tooling to accurately capture. Instead, we noted strength and dexterity for future evaluation, to be completed prior to public deployment.

The results of \textit{Collection Monitoring} led to three instances of additional, retargeted data collection efforts based on unexpected skew. During our initial data collection, there were multiple categories which did not match our previously described distributions.
We call out several in ~\autoref{fig:collection_comparison} to demonstrate the outcomes of our data collection re-targeting.
While all participants were employees from a large tech company, we did not anticipate our initial wave of data collection to be so skewed towards individuals between the ages of 21 and 40. In truth, we initially had no participants over 38. 
As a result, we emphasized diversity in participant age moving forward with some success---each consecutive wave improved the data coverage. Familiarity also directed our collection efforts, leading us to seek additional data for an intersectional subset of \texttt{Female} participants with \texttt{Large} phones and \texttt{Small} hands.

\begin{figure*}
  \centering
  \includegraphics[width=.9\linewidth]{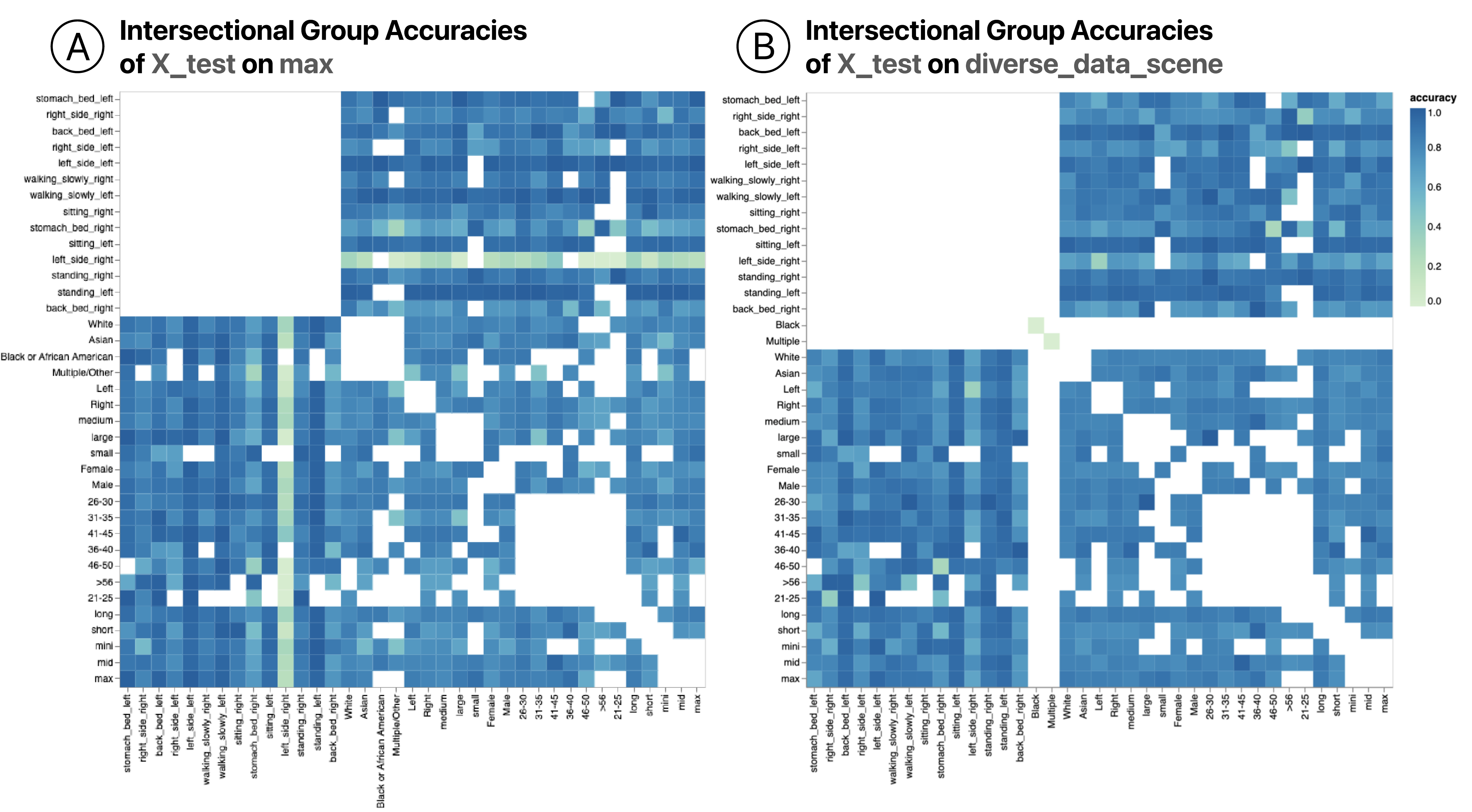}
  \caption{Comparison of intersectional groups of a less diverse model (A) to a diverse model (B). Striated accuracy across populations---as described by the metadata descriptions---performed worse when groups were left out, indicating that these characteristics were aligned with meaningful diversity in the data.}
  ~\label{fig:diversemodel}
\end{figure*}

\subsection{Collection Retargeting}
Similarly, we noted that the majority of our participants self-described as White, but that there were no Black or Indigenous participants whatsoever. Despite our best attempts, this was only partially amended; in optimizing across intersectional groups, we were not able to perfectly match our expected or updated distributions (at least within the context of this set of iterations). In contrast, our initially collected Handedness distribution perfectly matched our expectations. This prompted a discussion of whether this distribution was reasonable---despite matching the ratio of left versus right handedness in US populations, we believed that right and left handed individuals would type in dramatically different ways, thus may require sample size parity to be appropriately learned by the model. Examples of our retargeting efforts can be seen in ~\autoref{fig:collection_comparison}.
\section{Modeling}
Using Keras, our architecture included two 1D convolutional layers (standard for time series data), max-pooling layers, a dropout layer, a fully connected dense layer with ReLu activations, and a fully connected dense layer with softmax activations. We followed the generic \emph{Activity Recognition Chain}, which includes pre-processing, segmentation, feature extraction, and classification steps~\cite{cruciani2020feature} for our approach to modeling. We used 1D convolutional neural nets (CNNs)
for sequence classification.
The CNN's architecture performs feature extraction through the convolution of the input signal with a kernel (or filter).  To pre-process data, each session was segmented into $200ms$ windows, with $40ms$ overlap between segments. Session timing varied by how long it took participants to finish typing a given phrase. We corrected our IMU dataset to account for gravitational acceleration effects, then normalized (using a direct current blocker) and segmented IMU data in series to ensure all windows were of equal length. We discarded windows containing data from multiple sessions. For training, one sample equated to a window of the time series data. Training batch sizes were $256 \mbox{ (batch)} \times 200 \mbox{ (ms)} \times 3 \mbox{ (IMU)}$, where $ms$ is the window of time, $batch$ is the batch size, and IMU represents the three accelerometer data sources. All computations were run on NVIDIA V100 GPUs. Training all models took approximately 30 hours.

\section{Experiments}

\subsection{Diverse Data Experimental Protocol}
\label{dde_protocol}

We save every model, their weights, training accuracy and loss curves, and training dataset, keeping track of model version. 
Note that each model was trained 10 x 10 times, keeping the random seed stable for $r$ times, then repeating with a different seed. Weights from the models with highest accuracy across the trials were kept for later experiments. 

First, shuffle then randomly select train and test datasets such that no typing trials are split across datasets. We save the test dataset to evaluate every model trained with the current train/test split. Then, we compare the full dataset and train/test distributions using a visual check and earth mover's distance (EMD). If the difference in distributions are significant, repeat the first step. We group the data by category (\eg $Sex$) then by type (\eg $Female$). Next, randomly downsample each group such that each subset is of equal size. We compare the distribution of data sampled out to the downsampled group data, repeating sampling in the case of skew, then save the sampled out data. Lastly, we repeat the prior step but do so from the complete training set: this is the ``diverse'' dataset for the category. For each group within a category, we append all other groups together and then leave the current group out such that new training groups were $Female \bigcup Intersex$ (for example). For each set within a category, train a new model using our previously described 1D CNN.
We monitor for overfitting by setting early stopping based on loss with a patience of 4.

Finally, the influence of demographics on performance varied. In general, removing activity and age group subsets was more harmful to models than hand size, gender, or race subsets, but there were exceptions.

\begin{figure*}
  \centering
  \includegraphics[width=\linewidth]{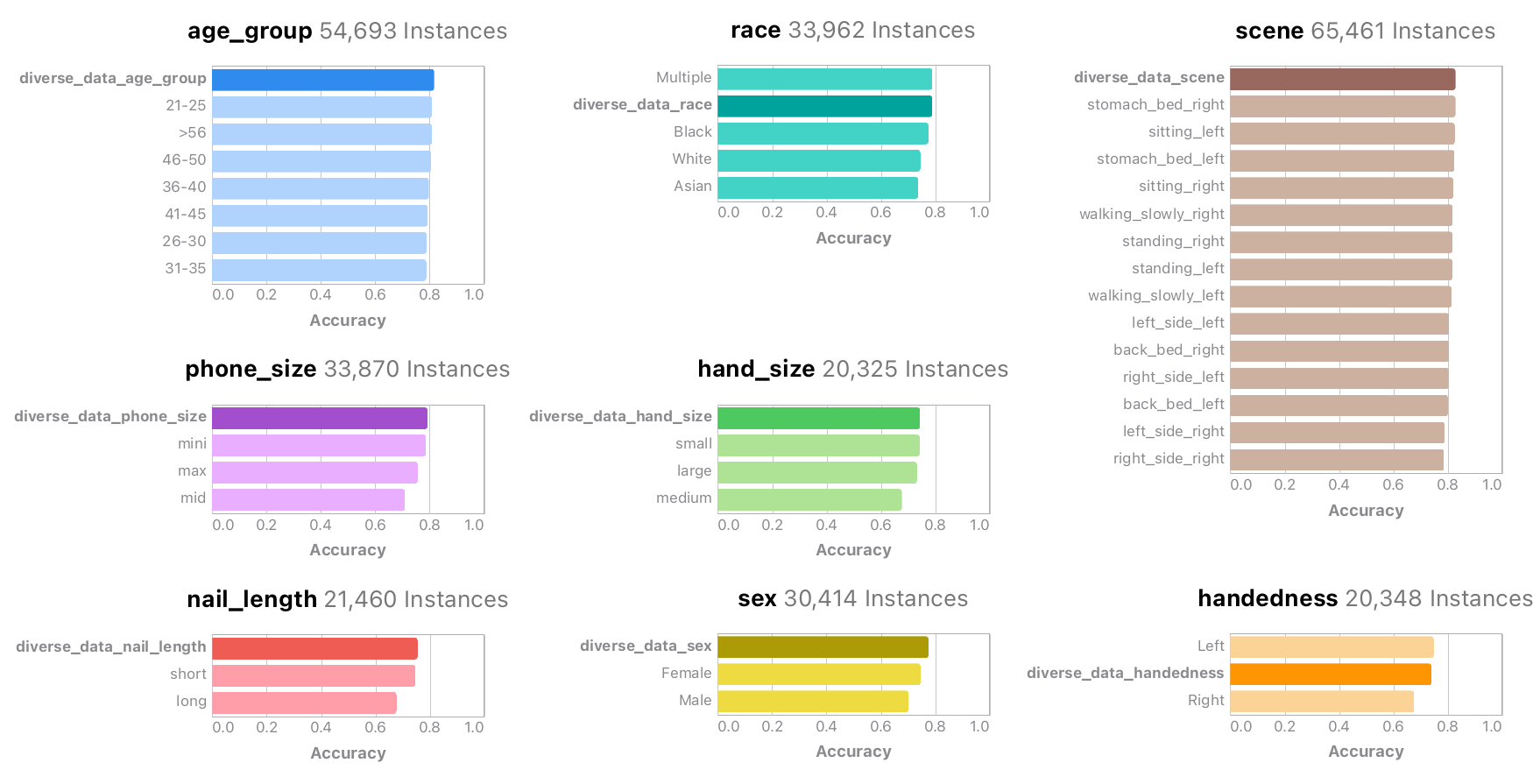}
  \caption{
    The performance of every model on the same test set, split across categories.
    For each category, a model's label indicates which subgroup was held out from its training set.
    To ensure fair comparison, within each category each model was trained on the same number of instances.
    Notice for each category that the ``diverse'' model (highlighted with a darker color), \ie the model with no subgroup held out, almost exclusively performs the best, despite having the same number of data instances as the other models.
  }
    \label{fig:best_trials}
\end{figure*}

\begin{figure*}
    \centering
    \includegraphics
[width=.7\linewidth]{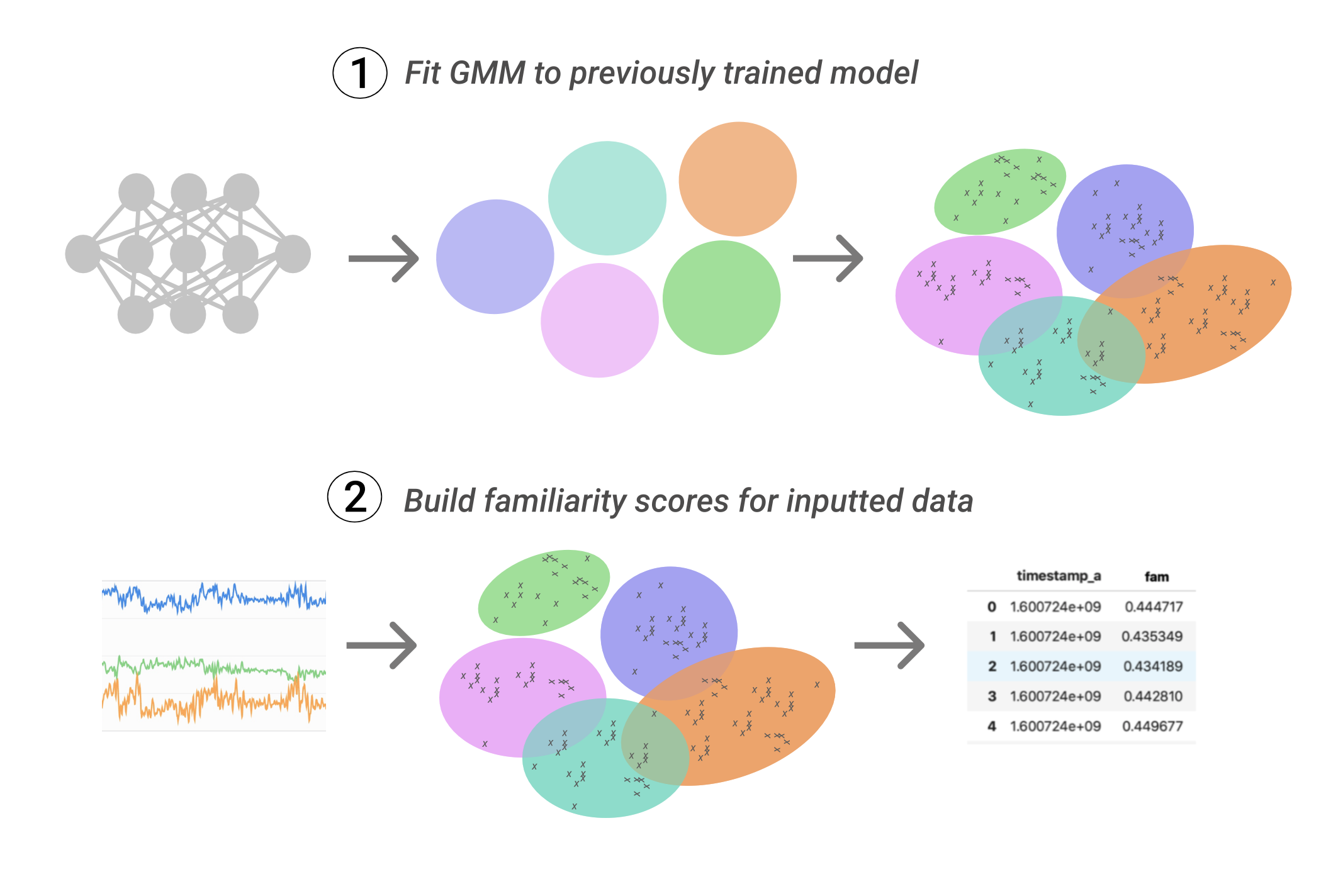}
    \caption{Overview of familiarity implementation. (1) We fit a Gaussian Mixture Model to our model of interest. (2) We present samples (e.g. our training data for ``self-familiarity'') to the GMM, returning log-likelihood scores.}
    \label{fig:gaussian}
\end{figure*}

\subsection{Familiarity Experiments: Capturing Noisy Data}
 Different layers of an NN capture distinct features of the input data~\cite{olah2018building}. Familiarity scores can therefore be extracted from any layer. Earlier layers capture fundamental structure found in the input data, while deeper layers capture semantic content. For this reason, we focus on the final dense layer, which holds the closest semantic alignment to human perception. A visual overview of the familiarity implementation can be seen in ~\autoref{fig:gaussian}.

In general, any DE technique can replace our implementation. Note that the same data shown to a different model is likely to obtain a different familiarity score; each sample is tightly coupled to how a specific model perceives it. This analysis cannot be done on the dataset without the guidance of an at least partially trained model.

\begin{figure*}
    \centering
    \includegraphics
    [width=.8\linewidth]{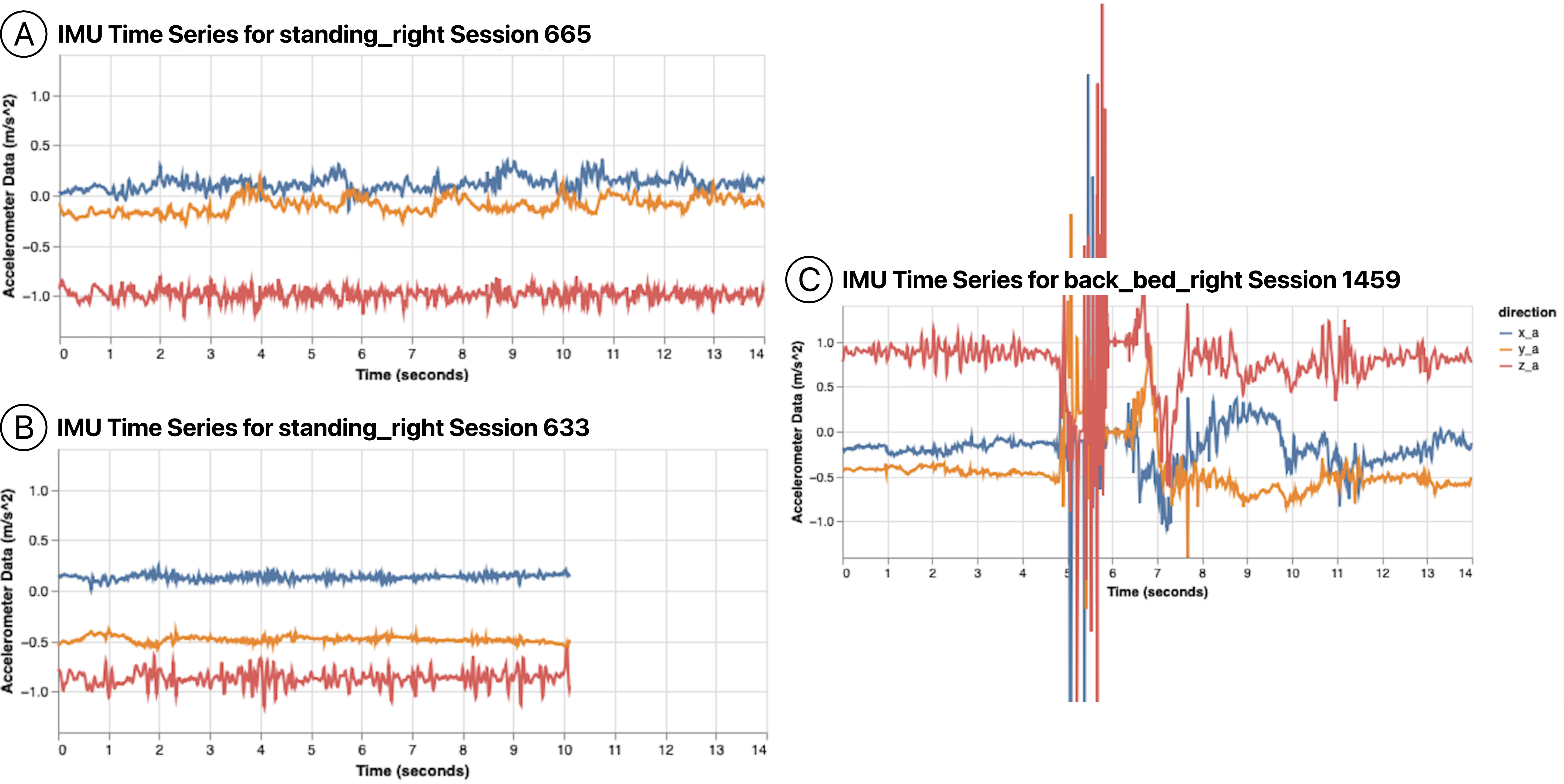}
    \caption{Distinct examples of data characterized as most ``unfamiliar'' to a model. (A) is an example of data we would consider out of distribution, (B) presents a case of sensor failure---the sensor stopped recording part way through the task---and (C) shows a particularly noisy sample, likely where someone dropped their phone mid-typing.}
\label{fig:noisyexamples}
\end{figure*}

A comparison of noisy data to OOD and more typical samples is shown in ~\autoref{fig:noisyexamples}.

\subsection{Diverse Data With Familiarity}

Given our intention was to \textit{encourage} diversity in our dataset, an ineffective sampling strategy might exacerbate edge case failures. 
We then select a range from the data distribution to sample the most familiar data from.
Thus we explored three general approaches:
\begin{enumerate}
    \item Replace $k$ most familiar samples with $k$ least familiar samples;
    \item Distributed sampling across a window of $k+i$ most familiar samples with $k$ least familiar samples, where $i$ represents a percentage of the overall training set
    \item Distributed sampling across a window of $k+i$ most familiar samples with $k+i$ least familiar samples, where $i$ is a percentage of the training set
\end{enumerate}

Each sampling mechanism was compared across multiple $k$ and $k+i$ values to determine the relative ``sweet spot'' for our sampling strategy given a particular training dataset. We randomly select $X$ percent of best and worst scores, varying percentage between 0.5\%---0.01\%. We train model(s) on each variation of window size and sampling percentage, repeating the previous steps $k$ times to ensure a multifold validation, then comparing the intersectional performances of $M_1$ to the new model $M_i$ trained on the familiarity-informed dataset. This is repeated per the scenario described in the full paper.

\begin{figure*}
    \centering
    \includegraphics
    [width=.6\linewidth]{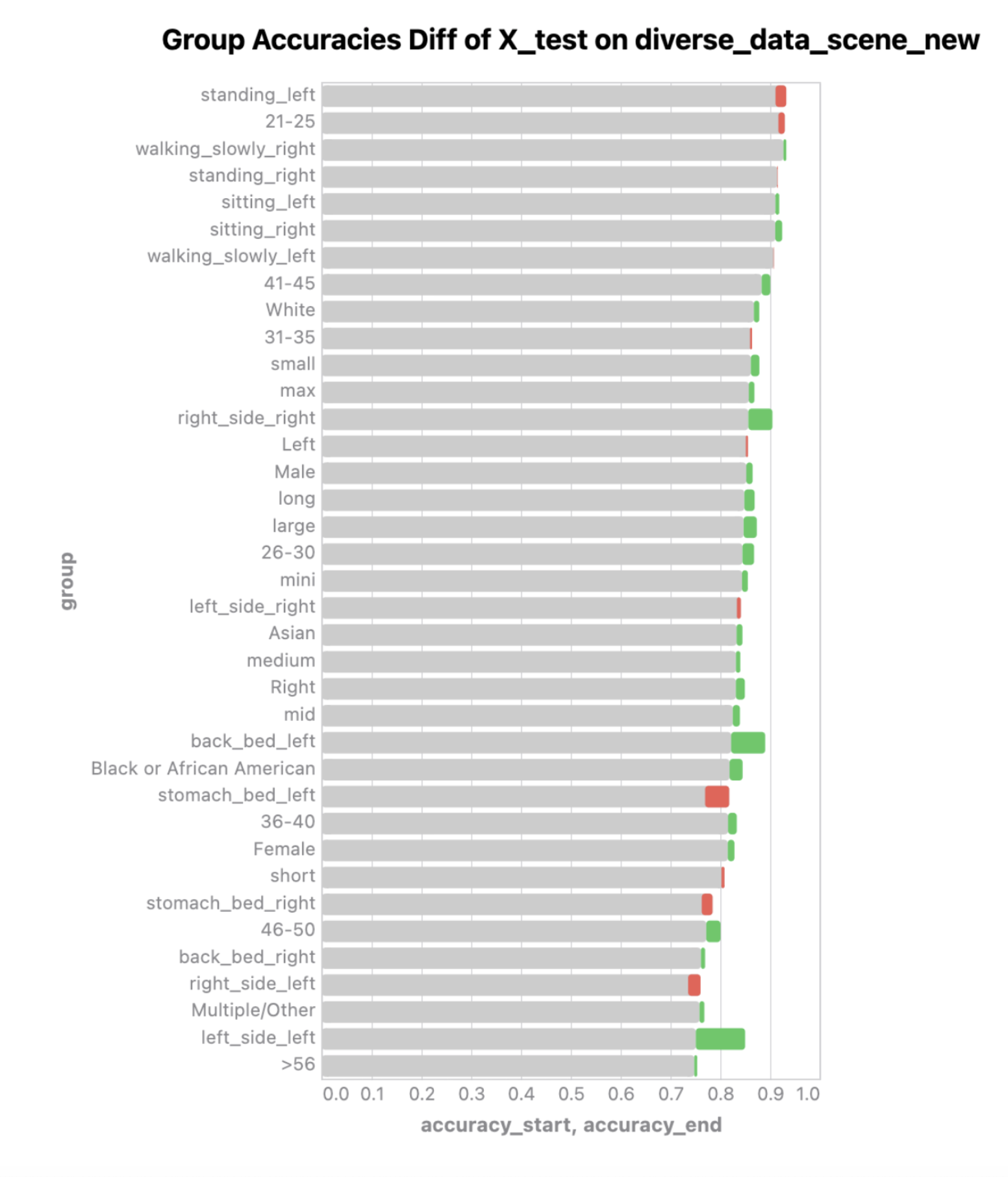}
    \caption{Comparison of accuracy on testing data before and after familiarity intervention. Minor regressions are shown, while accuracy improvements in contrast}
    \label{fig:fam_intervention}
\end{figure*}

\section{Limitations and Future Extensions of Familiarity}

\paragraph{Distinguishing between rare and noisy data}
A weakness of familiarity is that we have no current method of distinguishing noisy samples from out of distribution data. While an unfamiliar sample might stand out to the model, in many cases, human review is necessary to evaluate its implications. For this reason, \textit{in very noisy datasets}, familiarity may be a tool best used for debugging. Future research might seek to incorporate algorithmic methods of distinguishing sources of uncertainty, however there is currently little work on the topic. 
Existing research either relies on the learning rate as a proxy for discriminating types of uncertainty as aleatoric or epistemic~\cite{d2021tale}, for example, or builds on Bayesian networks~\cite{kendall2017uncertainties}. Both face various weaknesses, and there remains great need for additional techniques and evaluations.

\paragraph{Familiarity for new data}
While we explored familiarity largely from the perspective of ``self-familiarity''---that is, what the model has already been exposed to, it also introduces a mechanism by which we can understand how a model responds to data it's previously not seen. This may offer a mechanism of transparency through which future users could evaluate how a model responds to new data. In this work, we computed familiarity from a single layer. In future work, we will explore how familiarity computed at different layers can be leveraged. Given that each layer captures distinct features within the data, aggregating information across depths of the model may lead to more holistic identification of unfamiliar data \textit{and} what features are more specifically so to the model. One way to do this is through a Product of Experts (PoE) paradigm~\cite{hinton2002training} where each layer is considered an ``expert''. 

\paragraph{Comparisons against active learning}
On the surface, familiarity appears similar to active-learning (AL). AL requires practitioners to choose which data to use given a large collection. In our scenario, we must understand which data to \textit{collect} or gather when there is no additional data readily available to run the AL algorithm on. One way to circumvent this difference is to apply AL to the training set, and then extract statistics on the metadata that the AL algorithm indicates as most useful. For example, one could use the entropy of the logits:
high entropy on a data point might be an indication that the model is still uncertain about such type of data. An issue with such approach is that it implicitly assumes models are well calibrated, which is not always the case.

\paragraph{Interactive systems for designing data}
Designing data advances how we account for the interplay between data and model~\cite{hooker2021moving}, considering both within the deployment cycle to compensate for missteps in either.
In our case study, we use data visualizations (e.g., \autoref{fig:expected}) to compare practitioners' expectations against collected data distributions, then visualize familiarity to explore rare or noisy samples. 
The visualizations and interfaces used in this work are largely static; however, we see a great opportunity to build the designing data process into future interactive systems and tools for better data work and model evaluation.
From the HCI and visualization communities, there are a number of interactive systems that have helped ML practitioners explore their data~\cite{kyd, 2017facets, hohman2020understanding} and evaluate their models~\cite{amershi2015modeltracker,wexlerWhatIfToolInteractive2019, goertler2022neo}; for an in-depth survey on visual analytics for ML see~\cite{hohman2018visual}.
Directions for future interactive systems might include tools to help practitioners reflect on their data collection practices (e.g., digging into their expectations, as discussed in \Cref{sec:case-study}), or tools to direct familiarity analyses.

\section{Extended Literature Review}
\subsection{Fairness, Diversity, and Heterogeneity}

Fairness has become an increasingly common consideration for algorithmic contexts~\cite{drosou2017diversity}. However, the term has been conflated with justice, bias, ethical machine learning, equity, diversity, heterogeneity, and inclusion due to cross-disciplinary use and overloaded terminology~\cite{mitchell2020diversity, celis2018fair}. 

Algorithmic notions of fairness are often presented through mathematical formalizations intended to ensure anti-discrimination in the context of classification systems. They most commonly focus on ML model outcomes, but may also describe input data or how systems use model results~\cite{mitchell2020diversity}. ML models incorporate outcome-specific fairness by adding constraints to ensure either group or individual parity across classification error metrics~\cite{mitchell2020diversity}. Group fairness enhanced models~\cite{friedler2019comparative} use either anti-classification (wherein protected attributes are not included in decision making), classification parity (groups across protected attributes have similar predictive performance), or calibration (results are independent of protected attributes) methods~\cite{corbett2018measure}. In contrast to group fairness, individual fairness~\cite{dwork2012fairness} asks that individuals similar for a task be treated similarly throughout that task. While well-intentioned, each of these fairness enhancing approaches have incited criticism~\cite{corbett2018measure}.  

Mitchell \ea described diversity, heterogeneity, and fairness as related but distinct concepts~\cite{mitchell2020diversity}. In contrast to prior work by~\cite{asudeh2019assessing, zheng2012coverage}, which considers a more encompassing definition of diversity---a measure to capture nuance of collection based on variety of constituent elements---Mitchell \ea's diversity emphasizes attributes of social concern. 
They argue diversity measures that are not specific to social groups should instead be considered heterogeneity measures, though diversity has long been interchangeable with heterogeneity. Mitchell \ea's argument is understandably motivated by concerns of specificity and impact, but reconciling what data may confer biasing effects for social groups with what does not is difficult: can human-agnostic data exist in data collected by and in reference to people?

Fields with historic interest in diversity or heterogeneity include information retrieval, ecology, biology, organization science, sociology, and chemistry, all of which have developed or employed approaches to measure diversity. These approaches largely fall into one of the seven following buckets: geometric or distance-based, combinatorial, aggregate, utility and ranking, coverage, and hybrid distance and coverage based diversity measures~\cite{drosou2017diversity, zheng2012coverage}. Examples of geometric or distance-based measures of diversity include by dataset's volume~\cite{mouchet2010functional, chao2014unifying,anari2016monte,celis2016fair, whittaker1972evolution} or by variance such as in principle component analysis (PCA)~\cite{samadi2018price}. In general, these metrics more closely match with Mitchell \ea's definition of heterogeneity as they do not explicitly refer to features with societal import and context~\cite{mitchell2020diversity}. Ultimately, Mitchell~\ea's interpretation of diversity is intended to bring to light social inequalities found in ML products \emph{specifically}, while others emphasize the measure of heterogeneity---comprehensive variety within the dataset.

Diversity metrics in ML contexts have been used to direct bias mitigation efforts~\cite{celis2016fair}. Instances of this can be found in subset selection, where declarations of ``diversity constraints'' define expected frequencies for sensitive values that the data must satisfy~\cite{milani2020diversifying}, or to measure relative coverage within a dataset~\cite{mitchell2020diversity}. An alternative subset selection technique associates diversity scores to subsets then chooses subsets when probabilities are proportional to this score~\cite{celis2016fair, celis2018fair}. Uniformly, these methods act as stopgaps to biased and/or homogeneous data---particularly common in the context of data summarizations of big datasets~\cite{celis2018fair}. 

The ability to produce diverse subsets using diversity metrics does not guarantee fairness across samples in the form of appropriate representation of sensitive attributes~\cite{celis2018fair}. Partially, this is because fairness has multiple measures~\cite{mitchell2020diversity}. Fair treatment across social groups may also require \emph{different things for different contexts}. For instance, consider a dataset in which each data point has a gender. One notion of group fairness, useful for ensuring that the ground truth is not distorted, is proportional representation, \ie the distribution of sensitive characteristics in the output set should be identical to that of the input dataset. Another notion of fairness, argued to be necessary to reverse the effect of historical biases, could be equal representation where the representation of sensitive characteristics should be equal independent of the ratio in the input dataset. The plethora of metrics and evaluative tools vary and each have ongoing discussions of merit. Ultimately, datasets must be considered in relation to the use at hand, and the potential harm any failures may cause.

Qualitative research methods, statistics, and survey literature have historically managed representative data collection in a variety of ways, from expert panels to standards in population survey techniques, yet these methods face their own complications and do not necessarily translate to the needs of machine learning teams.

\subsection{Bias Mitigation}
While examples of algorithmic bias are often highlighted by media outlets, this frequency belies the difficulty of initial discovery---failures are hard to uncover during development, thus responsibility often falls to the public via product engagement. In the case of Google Photos, the model could not distinguish between a gorilla and a Black person and it was a member of the public that flagged the concerning labeling. Situations like racist photo identification algorithms are not uncommon, and once uncovered are responded to in a variety of ways. Google's response was to censor outputted labels such as ``chimpanzee'' and ``gorilla'' within the public Photos app search results~\cite{thevergegorilla}. In their case, the model could not distinguish between a gorilla and a Black person because it was not trained to do so---there were not enough instances of Black faces in the dataset~\cite{asudeh2019assessing}. Biased data and lack of diverse representation has broad impacts in domains beyond computer vision. Popular personalized voice agents, for example, struggle to recognize foreign accents under certain contexts such as understanding medication names when users \emph{correctly} pronounced them~\cite{palanica2019you}. 

Outside of censoring outputs, bias mitigation strategies can be divided into three stages of a ML model development: pre-training (\eg sample weighting or dataset balancing), in-training (\eg adding specific constraints in the function that is being optimized) and post-training (\eg by tweaking the prediction in order to ensure some fairness metric)~\cite{donini2018empirical}. Pre-training can be further divided into collection and post-collection. Despite the fundamental nature of data collection, typical technical approaches to bias mitigation focus on post-collection efforts---\eg modification of the dataset, reweighting and fine-tuning of hyperparameters, filtering output, or some combination therein---which largely act as stopgaps, are sensitive to underlying data~\cite{wieringa2020account} and ultimately may not resolve underlying issues at hand. 

Additional complications arise when working with data acquired independently, possibly through a process in which the data scientist has little or no control. This ``found data''~\cite{ramasamy2018recent} introduces unique challenges to ensuring data coverage for scientists and engineers. For both found data and big data contexts, post-collection approaches such as subset selection and class imbalance corrections like under- or over-sampling are introduced to counter bias and skew~\cite{japkowicz2002class}. Yet sampling methods can obfuscate information about appearance frequency in a dataset---if there are limited examples of X and so we oversample, then those instances of X are not unique and may cause the model to infer incorrect characteristics of a class, affecting accuracy metrics as well as production performance. These approaches are limited; for example, when improving a model without access to the original training dataset, balancing in the traditional sense is impossible. Regardless of the context, these data balancing approaches ignore data learnability: having an equal number of samples per class neglects the fact that some classes are inherently easier to learn than others~\cite{ben2019learnability, schapire1990strength, klawonn2019exploiting}.

Work on ML generalization has looked to avoid the issue of biases in data distribution and collection altogether by focusing on the causal relationships within data. Supposedly robust to   lack of coverage or variation within a dataset, Invariant Risk Minimization (IRM) relies on latent variables within the data---information not explicitly observed, but rather which are inferred from existing data---to learn concepts more closely aligned with ground truth~\cite{arjovsky2019invariant}. IRM has met been met with scepticism, however, and work by Rosenfeld~\ea found that IRM can ``fail catastrophically'' unless the test and training distributions are sufficiently similar~\cite{rosenfeld2020risks}, ultimately coming back to the intitial question of how to develop diverse datasets that more accurately reflect real world use cases.

\subsection{Contrasting Familiarity to Other Methods}
Lee et al \cite{lee2018simple} is most similar to our proposed method of familiarity. However, there are several differences between our work and theirs: they use class-conditional density estimation, while we do not use labels, thus fit one model across all classes. We use PCA to reduce activation dimensionality before fitting the model (DE in high dimensional space is difficult). Finally, we use a variational Bayesian estimation of the GMM; they estimate Gaussian mean and covariance differently.

\end{document}